\documentclass[aps,prd,nofootinbib,showpacs,superscriptaddress,preprint]{revtex4-1}
\pdfoutput=1
\usepackage[utf8]{inputenc}
\usepackage{amsmath,amssymb}
\usepackage{epsfig}
\usepackage{graphicx}
\usepackage{slashed}
\usepackage[usenames,dvipsnames]{color}
\usepackage[colorlinks,citecolor=blue]{hyperref}
\usepackage{soul}
\usepackage{color}
\usepackage{subfigure}
\begin{document}
%opening

\title{Gravitational origin of dark matter and Majorana neutrino mass with non-minimal  quartic inflation}

%%%%%%%%%   Authors   %%%%%%%%%%%%
\author{Debasish Borah}
\email{dborah@iitg.ac.in}
\affiliation{Department of Physics, Indian Institute of Technology Guwahati, Assam 781039, India}
\author{Suruj Jyoti Das }
\email{suruj@iitg.ac.in}
\affiliation{Department of Physics, Indian Institute of Technology Guwahati, Assam 781039, India}
\author{Abhijit Kumar Saha}
\email{aks@prl.res.in}
\affiliation{Theoretical Physics Division, Physical Research Laboratory, Navrangpura, Ahmedabad 380009, India}    
 
\begin{abstract}
We propose a minimal framework to address successful quartic inflation, dark matter (DM) production in the early universe and non-vanishing tiny Majorana neutrino mass from a common gravitational origin point of view. In this setup, the quartic inflation is revived successfully via non-minimal 
coupling of inflaton to gravity while the production of DM takes place from purely
gravitational effects through a misalignment mechanism. The generation of light Majorana neutrino mass is aided by 
explicit breaking of global lepton number symmetry through Planck suppressed operators involving non-zero vacuum expectation value (VEV) of the inflaton field. We present a detailed study of the DM yield in presence of non-minimal inflation,
considering both the metric and the Palatini formalisms of gravity wherever appropriate.
We reach at some interesting and different results in the DM sector compared to the earlier works in the similar direction with minimal inflationary background. Restricting to the light DM regime ($\mathcal{O}(1)$ keV- $\mathcal{O}(100)$ MeV) where classical production is expected to dominate over the quantum production, we numerically predict the DM mass by varying
the DM quartic and non-minimal coupling, to be consistent with relic density requirements. 
We also obtain some non-trivial dependence of DM phenomenology on some of the relevant parameters of the
inflation sector {\it e.g.} non minimal coupling and inflaton VEV. To explore the dependence on inflationary parameters further, we also estimate
the DM relic using the same mechanism for two other inflationary models consistent with latest
data and observe that one of these models predicts different range of DM mass upto hundreds of TeV.
\end{abstract}
\maketitle

%%%%%%%%%%%%%%%%%%%%%%%%%%%%%%%
\section{Introduction}
%%%%%%%%%%%%%%%%%%%%%%%%%%%%%%%%
There have been significant amount of evidences so far suggesting the presence of a non-luminous, non-baryonic form of matter, known as dark matter (DM) in the present universe \cite{Tanabashi:2018oca}. In particular, evidences from cosmology based experiments like Planck \cite{Aghanim:2018eyx} during last two decades and astrophysical observations over a much longer period of time \cite{Zwicky:1933gu, Rubin:1970zza, Clowe:2006eq, Salucci:2018hqu} suggest that a vast amount of matter in the present universe is dominated by DM. Planck 2018 data reveal that approximately $26\%$ of the present universe is composed of DM, which is about five times more than the ordinary luminous or baryonic matter. In terms of density parameter $\Omega_{\rm DM}$ and $h = \text{Hubble Parameter}/(100 \;\text{km} ~\text{s}^{-1} \text{Mpc}^{-1})$, the present DM abundance is conventionally reported as \cite{Aghanim:2018eyx}:
%\begin{equation}
$\Omega_{\text{DM}} h^2 = 0.120\pm 0.001$
%\label{dm_relic}
%\end{equation}
at 68\% CL. While none of the standard model (SM) particles can fit the requirements of a particle DM candidate, several beyond standard model (BSM) proposals have been put forward in the past few decades. Out of these, the weakly interacting massive particle (WIMP) is perhaps the most widely studied framework where a DM particle having mass and interactions typically around the electroweak scale can give rise to observed abundance after thermal freeze-out, a remarkable coincidence often referred to as the {\it WIMP Miracle} \cite{Kolb:1990vq,Arcadi:2017kky}.
 While typical interactions of WIMP with the SM particles are expected to show up in direct search experiments, no such signals have been observed yet, pushing the upper limit on DM-nucleon cross-section closer to the neutrino floor \cite{Aprile:2018dbl}. Non-observation of DM signal at such experiments have also increased the interest of particle physics community in looking for alternatives to WIMP paradigm. One such scenario which has got some attention recently is the non-thermal origin of DM \cite{Hall:2009bx}. For a recent review of such feebly interacting (or freeze-in) massive particle (FIMP) DM, please see \cite{Bernal:2017kxu}. 

%In this framework, DM candidate does not thermalise with the SM particles in the early universe due to its feeble interaction strength and the initial abundance of DM
%is assumed to be negligible. At some later stage, DM can be produced non thermally from decay or annihilation of other particles thermally present in the universe.

While motivations for WIMP as well as FIMP type of models are still there, one should also note that all the evidences in favour of DM so far are based on its gravitational interactions only. Therefore, it may sound natural to inquire the possibility of DM being produced in the early universe by virtue of a mechanism that relies on gravitational interactions only \cite{Ford:1986sy}. DM production through minimal coupling to gravity can occur during the transition to post-inflationary stage \cite{Chung:1998ua, Chung:1998zb, Kuzmin:1998uv, Kuzmin:1998kk} or during (p)reheating stage \cite{Greene:1997ge, Chung:1998rq, Garny:2015sjg}. Some recent works on vector DM production via minimal coupling to gravity can be found in \cite{Ahmed:2020fhc, Kolb:2020fwh}. Recently, production of superheavy scalar DM with masses as heavy as $10^{16}$ GeV was discussed in the context of a gravitational misalignment mechanism \cite{Babichev:2020xeg}. In this process the scalar DM field being $Z_2$ symmetric couples non-minimally to gravity. This particular coupling induces a time dependent spontaneous $Z_2$ breaking expectation value at some earlier time. As the universe expands, the expectation value decreases and at some point the $Z_2$ symmetry gets restored following which the DM field starts to oscillate around zero with an initial amplitude depending on the inflationary background as well as the relevant parameters of the DM sector. 
%Such time dependent spontaneous symmetry breaking potential of the DM scalar field was generated via non-minimal coupling to gravity or the Ricci scalar. 
One similar scenario with Ricci curvature induced symmetry breaking, while focusing on the light DM mass regime, was discussed recently by the authors of \cite{Laulumaa:2020pqi}. Such non-minimal coupling naturally arises for quantum fields in curved spacetime \cite{Callan:1970ze, Tagirov:1972vv} and has also received lots of attention in the context of inflation \cite{Bezrukov:2007ep, Bezrukov:2010jz, Kaiser:2013sna, Schutz:2013fua, Pallis:2014cda,Gumjudpai:2016ioy,Tenkanen:2017jih, Shokri:2019rfi, Takahashi:2020car, vandeVis:2020qcp, Borah:2020wyc} as well as reheating \cite{Figueroa:2016dsc, Dimopoulos:2018wfg, Opferkuch:2019zbd, Bettoni:2018utf, Bettoni:2018pbl, Bettoni:2019dcw}. The same non-minimal coupling to gravity can also lead to DM production, as discussed already in several works including \cite{Alonso-Alvarez:2018tus, AlonsoAlvarez:2019cgw, Markkanen:2015xuw, Fairbairn:2018bsw, Ema:2016hlw, Ema:2018ucl, Velazquez:2019mpj, Chung:2018ayg}.

In this paper we revisit the DM production through the gravitational misalignment process in a non-minimal quartic inflationary background, discussed earlier. In particular, to incorporate the light Majorana neutrino mass simultaneously, we develop a concrete particle physics model in which the SM is extended with right handed neutrinos and two other real singlet scalar fields to play the role of DM and inflaton respectively. We first review the non-minimal quartic chaotic inflation in view of latest Planck experimental data \cite{Akrami:2018odb} and find out the allowed parameter space. Interestingly, the inflaton having non zero VEV is responsible to yield light neutrino masses by seesaw mechanism through a non-renormalisable lepton number violating operator. Such lepton number violation at Planck scale may arise due to quantum gravity effects \cite{Borah:2013mqa, Davoudiasl:2020opf, Ibarra:2018dib, Borah:2019ldn}. More importantly, we have analysed the DM production for both the Palatini and metric formalisms of gravity by considering suitable magnitudes of the non-minimal coupling of inflaton. Appropriate conformal factors are also taken into account while solving the simultaneous Boltzmann equations involving inflaton, DM and SM radiation. We also restrict ourselves to the light DM mass regime where it is anticipated that the classical production of DM dominates over the quantum production. It is to be noted that similar kind of DM production has been worked out by authors of \cite{Babichev:2020xeg, Laulumaa:2020pqi} however considering minimal inflationary scenario. Although we have an enlarged framework with broader range of motivations, we like to report some important implications of the non-minimal inflation on the dynamics of DM phenomenology compared to earlier works \cite{Babichev:2020xeg, Laulumaa:2020pqi}. To be specific, it is observed that the DM relic density carries some distinct dependencies on the parameters of the non-minimal inflation sector, not observed in earlier works. In each of the Palatini and metric cases, we numerically estimate the DM mass to be consistent with relic density requirements as function of DM quartic and non-minimal couplings which turn different compared to the cases with minimal inflationary background.
Additionally, the VEV of the inflaton field responsible for right handed neutrino mass also leaves some imprints on the DM sector. In continuation, to investigate the parameter dependence of DM phenomenology on inflation sector further, we generalise the study to two other inflationary models namely hilltop and natural inflation and found different DM mass range in one of these cases. 

%with masses from a few keV to a few hundreds of MeV can be consistent with relic density requirements depending on the DM quartic and non-minimal couplings. 

This paper is organised as follows. In section \ref{sec1}, we discuss our model where both DM and inflaton fields have non-minimal coupling to gravity. In section \ref{sec2}, we discuss inflationary dynamics and predictions of the model followed by the details of gravitational DM production in section \ref{sec3}. We extend our study to two other inflationary frameworks in section \ref{sec4} and finally conclude in section \ref{sec5}.

%%%%%%%%%%%%%%%%%%%%%%%%%%%%%%%%
\section{The Model}
\label{sec1}
%%%%%%%%%%%%%%%%%%%%%%%%%%%%%%%%
We consider a minimal extension of the SM by two real scalar fields ($S$ and $\chi$) and three Majorana right handed neutrinos, all being singlet under the SM gauge symmetries. We also impose a global $U(1)_L$ symmetry which remains conserved at the renormalisable level, just like in the SM. While the two singlet scalars can give rise to the desired inflation and DM phenomenology via non-minimal coupling to gravity, the right handed neutrinos can give rise to non-zero neutrino masses, as confirmed by neutrino oscillation experiments \cite{Tanabashi:2018oca}. The particle content of our model and their transformations under the global lepton number symmetry imposed are shown in table \ref{tab1}.
\begin{table}[h]
\begin{center}
  \begin{tabular}{ | c | c | c | c | }
    \hline
    Fields &~ $\chi$~ &~ $S$~ &~ $N$~ \\ \hline
    $U(1)_{L}$ & $0$ & 0 & 1\\ 
    $ SU(3)_c$ & 1 & 1 & 1 \\
    $ SU(2)_L$ & 1 & 1 & 1 \\
    $ U(1)_Y$ & 0 & 0 & 0 \\
    
    \hline
  \end{tabular}
  \end{center}
  \caption{New particle content of the model and their transformations under the symmetries of the model.}
    \label{tab1}
  \end{table}

While a conserved lepton number symmetry can lead to Dirac neutrino masses with appropriate fine-tuning of Dirac Yukawa couplings, we consider a scenario where global lepton number is broken by Planck scale suppressed operators. This is also supported by the argument that any generic theories of quantum gravity should not respect discrete as well as continuous global symmetries \cite{Abbott:1989jw, Kallosh:1995hi, Hawking:1974sw}. Consequences of such global symmetry breaking by quantum gravity effects on neutrino mass and mixing have been explored in several works \cite{Borah:2013mqa, Davoudiasl:2020opf, Ibarra:2018dib, Borah:2019ldn} \footnote{See \cite{Dvali:2016uhn, Barenboim:2019fmj} for other studies related to gravitational origin of light neutrino masses.}. While the Planck scale suppressed Weinberg operator involving SM Higgs and lepton doublets \cite{Weinberg:1979sa} will give a tiny sub-dominant contribution to Majorana light neutrino masses, we can write a Planck suppressed dimension five operator involving the inflaton field and right handed neutrinos which, after inflaton field acquires a non-zero vacuum expectation value (VEV) can give rise to a Majorana mass term, leading to the implementation of type I seesaw mechanism.

The relevant part of the Lagrangian can be written in three different parts as follows.
\begin{align}
&-\mathcal{L}_1\supset \frac{\lambda_S}{4}\left(S^2-v_S^2\right)^2+\frac{\xi_S}{2} S^2 R+ \frac{\lambda_{SH}}{2}S^2 (H^\dagger H)\\
&-\mathcal{L}_2\supset \frac{1}{2}m_\chi^2\chi^2+\frac{\lambda_\chi}{4}\chi^4-\frac{\xi_\chi}{2}\chi^2 R+ \frac{\lambda_{\chi H}}{2}\chi^2 (H^\dagger H)+\frac{\lambda_{\chi\phi}}{4}\chi^2 S^2\\
&-\mathcal{L}_3=Y_\nu\overline{l_L}\tilde{H} N+\frac{Y_N}{2M_P}~SS N N+ {\rm h.c.}
\end{align}
Here $H$ denotes the SM Higgs doublet. In the Lagrangian we do not write the odd power terms for the fields $S$ and $\chi$. This could be easily realised by considering presence of $Z_2\times Z_2^\prime$ discrete symmetries under which $S$ and $\chi$ field transform non-trivially respectively. Appropriate UV completion of such scenarios can provide an origin of such discrete symmetries which we do not pursue for the sake of minimality. Although we have written quartic interaction terms of DM field $\chi$ with the SM Higgs as well as the inflaton field which are allowed by the symmetries, we will ignore them in our analysis as required by the criteria of purely gravitational origin of DM.

After the electroweak symmetry breaking the VEV's of the scalar fields can be denoted as
\begin{align}
\langle H^\dagger H\rangle= \frac{v^2}{2}~~~~~~~\langle S\rangle =v_S.
\end{align}
The non-zero VEV of S field ($v_S$) assists in generating the Majorana mass for the right handed neutrinos as $M_R=\frac{Y_N v_{S}^2}{M_P}$. With this the light neutrino mass can be obtained via type I seesaw mechanism as
\begin{align} 
M_\nu=m_D^T M_R^{-1} m_D,
\end{align}
where $m_D=Y_\nu v/\sqrt{2}$ is the Dirac mass matrix assuming $m_D \ll M_R$. One can quantify the Yukawa coupling matrix $Y_\nu$ by the use of Casas-Ibarra parametrisation \cite{Casas:2001sr}
\begin{align}
Y_\nu=\frac{\sqrt{2}}{v}\sqrt{M_N} \mathcal{R} \sqrt{m_\nu^d}~U^\dagger_{\rm PMNS},\label{eq:casas1}
\end{align}
where $m_\nu^d, M_N$ are the light and heavy neutrino mass matrices in the diagonal basis and $\mathcal{R}$ is an arbitrary complex orthogonal rotational matrix with three mixing angles.
The $U_{\rm PMNS}$ is the Pontecorvo-Maki-Nakagawa-Sakata (PMNS) leptonic mixing matrix that diagonalises the light neutrino mass matrix in the limit of diagonal charged lepton mass matrix and carries all the information of neutrino mass and mixing angles. A general parametric form of $U_{\rm PMNS}$ is given by,
\begin{align}
U_{\rm PMNS}=\begin{pmatrix}
c_{12}c_{13} & s_{12} c_{13} & s_{13}e^{-i\delta}\\
-s_{12}c_{23}-c_{12}s_{23}s_{13}e^{i\delta} & c_{12}c_{23}-s_{12}s_{23}s_{13}e^{i\delta} & s_{23}c_{13}\\
s_{12}s_{23}-c_{12}c_{23}s_{13}e^{i\delta} & -c_{12}s_{23}-s_{12}c_{23}s_{13}e^{i\delta} & c_{23}c_{13}
\end{pmatrix} U_{\rm Maj},\label{eq:PMNS}
\end{align}
where $c_{ij}=\cos\theta_{ij}$ and $s_{ij}\sin\theta_{ij}$. The phase $\delta$ stands for the Dirac CP phase. The $U_{\rm Maj}$ is a diagonal matrix expressed as diag($1,~e^{i\alpha},~e^{i\beta}$) where $\alpha$ and $\beta$ are the Majorana CP phases. 
%Combining eqn. \ref{eq:casas1} and eqn. \ref{eq:PMNS} it is trivial to estimate the $m_D$ matrix by satisfying the allowed ranges of neutrino mass and mixing angles.
%%%%%%%%%%%%%%%%%%%%%%%%%%%%%%%%%%%%%%%%%%%%%%%%%%%%%%%%%%%%%%%%%%%%%%
\section{Inflation}
\label{sec2}
%%%%%%%%%%%%%%%%%%%%%%%%%%%%%%%%%%%%%%%%%%%%%%%%%%%%%%%%%%%%%%%%%%%%%%
In this section, we describe the dynamics of inflation in detail and its predictions in view of the present experimental bounds. As pointed out earlier, we identify the scalar field $S$ as the inflaton field. Its action in the Jordan frame takes the following form \cite{Bostan:2018evz}
\begin{align}
 S_J=\int d^4x\sqrt{-g}\Bigg[\frac{M_P^2}{2}~\Omega(S)^2~R-\frac{1}{2}g^{\mu\nu}(\partial_{\mu}S)(\partial_{\nu}S)-V_J(S)\Bigg],
\end{align}
where $\Omega(S)^2=1+ \frac{{\xi_S} (S^2-v_s^2)}{M_P^2}$, $V_J=\frac{\lambda_S}{4} (S^2-v_S^2)^2$ and $g^{\mu\nu}$ is the space-time metric in the $(-,+,+,+)$ convention. During inflation, we can ignore the contribution of sub-Planckian $v_S$ in the action with respect to the super-Planckian inflaton field values.

Now, in an ideal situation, to work with pure canonical field theory in Einstein frame, we need to get rid of the non-minimal couplings of both $S$ and $\chi$ and correspondingly bring their kinetic terms into canonical form. However, as noted in \cite{Kaiser:2010ps}, this is not possible in principle to make a theory completely canonical (including the gravitational sector) in the presence of two or multiple 
non-minimal couplings to gravity. Alternatively, we can follow a sort of different route. Since the dark matter field is a spectator field and always remains sub-Planckian with amplitude $\chi(t)\ll M_P$, it affects the Einstein's equations negligibly (in contrast to non-minimal inflaton field) and we can still work in the regime of canonical inflaton field only. Indeed, authors of \cite{Alonso-Alvarez:2018tus} have argued that the dynamics of dark matter production
in presence of dark matter- gravity non-minimal coupling is more or less equivalent while analysing in the Einstein or Jordan frame, considering sub-Planckian DM field. With this notion, we pursue our analysis in a frame whis is not purely Einstein frame but rather can be assumed as an intermediate frame \cite{Gialamas:2020snr, Gialamas:2021enw}. For reader's convenience we term it as a ``new'' frame here onwards.

We then use the following conformal transformation in order to remove the inflaton-gravity non-minimal coupling so that we can use the standard description of calculating inflationary observables
\cite{Capozziello:1996xg, Kaiser:2010ps}:
\begin{equation}
\hat{g}_{\mu\nu}=\Omega^{2}g_{\mu\nu},~~\sqrt{-\hat{g}}=\Omega^{4}\sqrt{-g},\label{conftrans}
\end{equation}
 %In the above transformation, $\hat{g}$ represents metric in the Einstein frame.
 Further, to make the kinetic term of the inflaton canonical, we redefine $S$ by \cite{Tenkanen:2017jih}
\begin{equation}
\frac{d\phi}{dS}=\sqrt{\frac{\Omega(S)^{2}+\frac{6f\xi_S^{2}S^{2}}{M_{P}^{2}}}{\Omega(S)^{4}}},
\label{fieldtrans}
\end{equation}
where $\phi$ is the canonical field, and $f$ is a parameter which equates to $1$ in the metric formalism. The case of $f=0$ is known as Palatini formulation of gravity\footnote{See \cite{Bauer:2010jg, Rasanen:2017ivk, Jinno:2019und} for Higgs inflation in Palatini formalism. Gravitational DM production in Palatini formalism for superheavy DM mass have been recently studied in \cite{Karam:2020rpa}.}. In metric formalism, the $R_{\mu\nu}(\Gamma)$, where $\Gamma$ is the connection, is function of the metric $g_{\mu\nu}$. On the other hand, in the Palatini formalism, the connection $\Gamma$ and the metric $g_{\mu\nu}$ are treated independently. 

After conformal transformation the potential for inflaton in the new frame in terms of the canonical field $\phi$ is given by,
\begin{equation}
V_{E}(\phi)=\frac{V_{J}\big(S(\phi)\big)}{\big [\Omega\big(S(\phi)\big)\big ]^{4}}.
\label{pottrans}
\end{equation}
Below we discuss the inflationary predictions of this model for two cases mentioned above namely, (I) $f=0$ and (II) $f=1$ separately.

\noindent \textbf{Case I:} In Palatini gravity with $f=0$, the transformation given by equation \eqref{fieldtrans} allows us to analytically solve for $S=S(\phi)$ as given by,
\begin{align}
 S(\phi)=&\frac{e^{-\frac{\sqrt{\xi_S} \phi }{M_P}}}{\xi_S  \left(2 M_P^2-2 \xi_S  v_S^2\right)} \Bigg[M_P^3 \sqrt{\xi_S} e^{\frac{2 \sqrt{\xi_S } \phi }{M_P}}-M_P^3 \sqrt{\xi_S}+M_P^2 \xi_S  v_S e^{\frac{2 \sqrt{\xi_S } \phi }{M_P}}+M_P^2 \xi_S  v_S\nonumber\\
 &~~~~~~~~~~~~~~~~~~~~~~~~~~~~\xi_S ^2 v_S^3 e^{\frac{2 \sqrt{\xi_S } \phi }{M_P}}-M_P \xi_S ^{3/2} v_S^2 e^{\frac{2 \sqrt{\xi_S} \phi }{M_P}}+M_P \xi_S^{3/2} v_S^2-\xi_S^2 v_S^3\Bigg].
 \label{eq:PalatiniTransO}
\end{align}
This, in the limit $S\gg v_S$ during inflation can be simplified to,
\begin{equation}
S=\frac{M_P}{\sqrt{\xi_S}}\sinh\big(\frac{\sqrt{\xi_S}\phi}{M_P}\big).
\end{equation}
This simplifies the inflation potential in the new frame (equation \eqref{pottrans}) as function of the canonical field $\phi$ only:
\begin{equation}
V(\phi)=\frac{\lambda}{4\xi_S^{2}}M_{P}^{4}\frac{\sinh^{4}\big(\frac{\sqrt{\xi_{S}}\phi}{M_{P}}\big)}{\bigg(1+\sinh^{2}\big(\frac{\sqrt{\xi_{S}}\phi}{M_{P}}\big)\bigg)^{2}}.
\label{eq:5}
\end{equation}
Next we proceed to find out the inflationary predictions in view of the most recent Planck results \cite{Akrami:2018odb}.
 
 In the slow-roll inflation, the standard parameters are $\epsilon$ and $\eta$ which are defined by:
 \begin{align}
 \epsilon(\phi)=\frac{M_P^2}{2}\left(\frac{V^\prime(\phi)}{V(\phi)}\right)^2 ~~~~~~~\eta(\phi)=M_P^2\left(\frac{V^{\prime\prime}(\phi)}{V(\phi)}\right).
 \end{align}
 With this in the slow-roll approximation, the observables of the inflation are parametrised as:
 \begin{align}
 &n_s=1-6\epsilon+2\eta,\\
 & r=16\epsilon,\\
 &P_s=\frac{V}{24 M_P^4\pi^2 \epsilon},
 \end{align}
 where $r$, $n_s$ and $P_s$, in their standard notations, are known as tensor to scalar ratio, spectral index and curvature perturbation spectrum respectively. These parameters have to be determined at horizon exit of the inflaton ($\phi_*$) during the onset of inflation. The number of e-folds is defined by:
 \begin{align}
 N_e=\frac{1}{M_P}\int_{\phi_{\rm end}}^{\phi_*}\frac{V(\phi)}{V^\prime(\phi)}d\phi,
 \end{align} 
 where $\phi_{\rm end}$ stands for the end of inflation.

\begin{figure}[h]
\noindent \begin{centering}
\includegraphics[height=7cm,width=8cm]{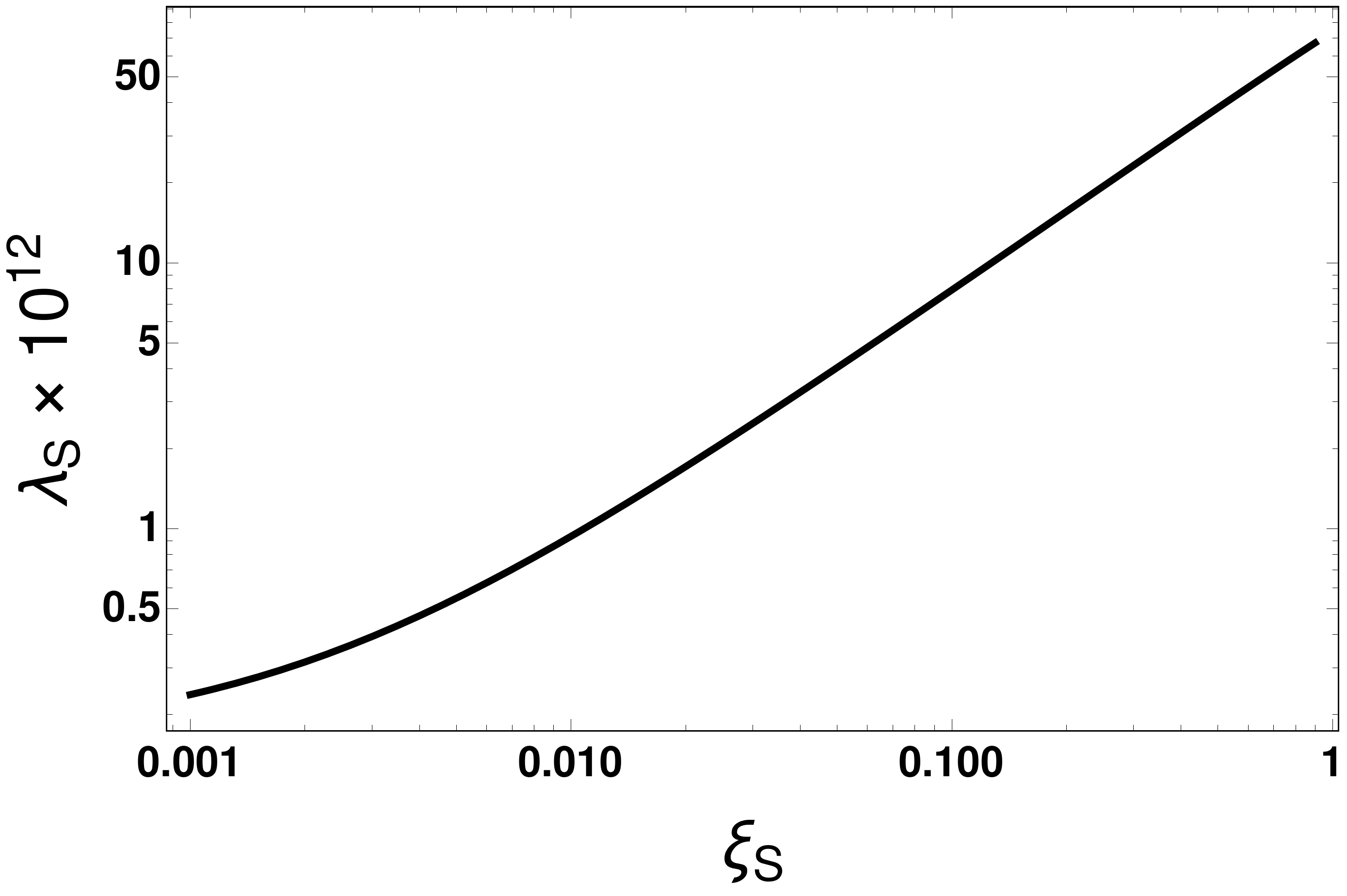}~~~~
\includegraphics[height=7.2cm,width=8.3cm]{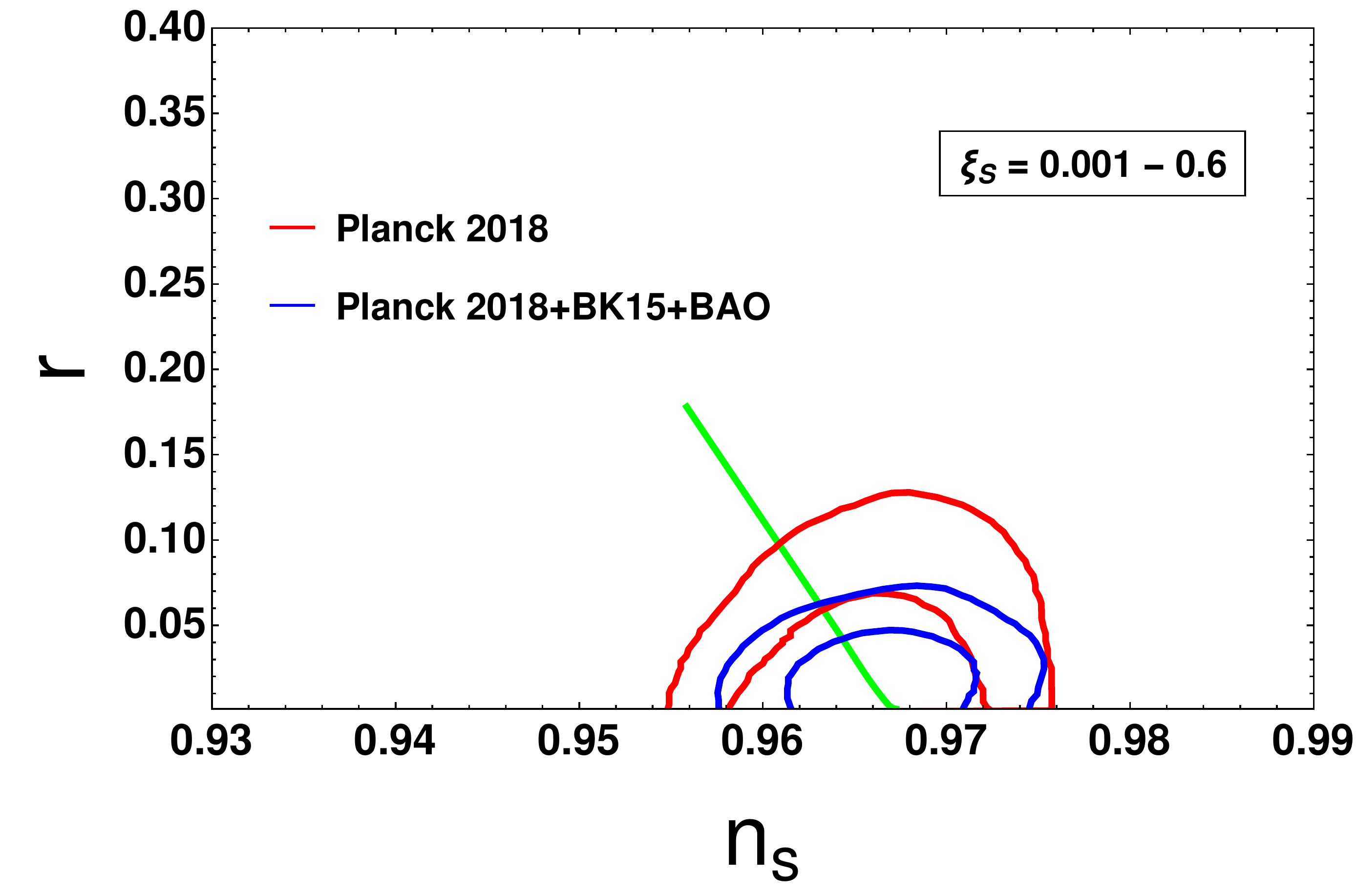}
\par\end{centering}
\caption{\label{fig:Inflation}Case I: (Left panel) Variation of $\lambda_S$ as a function $\xi_S$ in order to produce the correct order of curvature
perturbation spectrum $P_S$. (Right panel) Predicted values of $n_s$ and $r$ along with the $1\sigma$ and $2\sigma$ bounds from Planck 2018+BAO+BK15 considering $N_e=60$.}
\end{figure}

In the left panel of Fig. \ref{fig:Inflation}, we show a contour line for the observed value of $P_s=2.2\times 10^{-9}$ in $\lambda_S-\xi_S$ plane. In the right panel we present the model prediction for $n_s-r$ values by varying $\xi_S$ against the Planck $1\sigma$ and $2\sigma$ bounds considering number of e-folds as 60.  For comparison purpose we also insert the Planck 2018+BK15+BAO 1$\sigma$ and $2\sigma$ bounds \cite{Akrami:2018odb}. It is evident that the present setup is able to provide set of $n_s-r$ values (green solid line), consistent with the experimental constraints.

\vspace{2mm}
\noindent \textbf{Case II:} We next consider the metric case, i.e., $f=1$ (see equation \eqref{fieldtrans}). For $\xi_S\lesssim \mathcal{O}(1)$, it is non-trivial to find the analytical behaviour of $S$ as function of $\phi$. Hence we work in $\xi_S\gg 1$ regime where the potential resembles the well known form of Starobinsky like inflation. The field transformation of $S$ and the corresponding potential $V(\phi)$ are read as: 
\begin{align}
&S=\frac{M_{P}}{\sqrt{\xi_S}}\sqrt{\exp\bigg(\sqrt{\frac{2}{3}}\frac{\phi}{M_{P}}\bigg)-1}        ,\\
&V(\phi)=\frac{\lambda_S M_{P}^{4}}{4\xi_{S}^{2}}\bigg[1-\exp\left(-\sqrt{2/3}\frac{\phi}{M_{P}}\right)\bigg]^{2}. \label{Starpot}
\end{align}  
Considering $N_e=60$, we find $r\sim 0.003, n_s\sim 0.967$, which are well inside the Planck 2018 $1\sigma$ bounds. This set of values is insensitive to the variation of $\xi_S$. In order to produce correct value of scalar perturbation spectrum, the following condition needs to be satisfied:
\begin{align}
\xi_S\sim49000\sqrt{\lambda_S}.\label{PSreltn} 
\end{align}

\begin{table}
\begin{centering}
\begin{tabular}{|c|c|c|c|c|c|c|}
\hline 
Sl no & ~$\xi_{S}$~ & ~$\lambda_{S}$~ &~ $\kappa$ &~$r$~ & ~$n_{s}$  & ~$\Gamma$(GeV)~\\
\hline 
CI-1 & $0.05$ & $5.42\times10^{-12}$ & $10^{-13}$  & $0.0159$ & $0.972$ & $7.32\times10^{10}$\\ 
\hline 
CI-2 & $0.1$ & $9.5\times10^{-11}$ & $6\times10^{-14}$  & $0.0072$ & $0.9711$ & $3.32\times10^{10}$\\
\hline 
CI-3 & $0.3$ & $2.5\times10^{-11}$ & $3\times10^{-14}$  & $0.0021$ & $0.9695$& $1.02\times10^{10}$\\
\hline 
CI-4 & $0.5$ & $4.38\times10^{-11}$ & $3.9\times 10^{-14}$  & $0.0013$ & $0.9702$& $1.02\times10^{10}$\\
\hline 
\end{tabular}
\end{centering}
\caption{Case I: Predictions of inflationary observables for non-negligible $\kappa$. We also show the decay width of inflaton for each of the benchmark points. The value of $v_S$ is considered to be 0.08$M_P$.}
\label{tab:tabC1}
\end{table}

So far, we have ignored the radiative corrections to the inflationary potential considering $\lambda_{SH}$ to be negligible. The radiatively corrected inflationary potential is given by (during inflation in the new frame) \cite{Okada:2010jf,Bostan:2019fvk,Bostan:2019uvv}:
\begin{align}
 V_N(S(\phi))\simeq \frac{\frac{\lambda_S(\phi)}{4}S(\phi)^4}{\left(1+\frac{\xi_S S(\phi)^2}{M_P^2}\right)^2}+\frac{\kappa S(\phi)^4{\rm Log}\left[\frac{S(\phi)}{\mu}\right]}{\left(1+\frac{\xi_S S(\phi)^2}{M_P^2}\right)^2},
\end{align}
where $\kappa\simeq \frac{\lambda_{SH}^2}{32 \pi^2}$ and $\mu$ is the renormalistaion scale which can be identified with the cut-off scale $\sim M_P/\xi_S$ \cite{Okada:2010jf}. The second term indicates the contribution to the inflationary potential due to Higgs mediated loop diagram. For sufficiently larger $\kappa$ value, this could bring some amount of deformation which can alter the inflationary observables found earlier. In table \ref{tab:tabC1} and table \ref{tab:tabC2}, we mention a few benchmark points and the corresponding magnitudes of inflationary observables in presence of non-zero $\kappa$ considering the Palatini and metric cases respectively. All the benchmark points for both Palatini and metric cases are allowed by the observational constraints from Planck 2018+BAO+BK15 data.
\begin{table}
\begin{centering}
\begin{tabular}{|c|c|c|c|c|c|c|}
\hline 
Sl no & ~$\xi_{S}$~ & $~\lambda_{S}$~ &~ $\kappa$ ~ &~ $r$~ & ~$n_{s}~$ & ~$\Gamma$~(GeV)\\ 
\hline 
CII-1 & $150$ & $1\times10^{-5}$ & $2\times10^{-8}$  & $0.0035$ & $0.9703$& $1.35\times10^{11}$\\
\hline 
CII-2 & $175$ & $1.41\times10^{-5}$ & $3.32\times10^{-8}$ & $0.0036$ & $0.9706$ & $1.882\times10^{11}$\\
\hline 
CII-3 & $200$ & $1.9\times10^{-5}$ & $6.32\times10^{-8}$  & $0.0038$ & $0.9717$ & $3.09\times10^{11}$\\
\hline 
\end{tabular}
\end{centering}
\caption{Case II: Predictions of inflationary observables for non-negligible $\kappa$. We also show the decay width of inflaton for each of the benchmark points. The value of $v_S$ is considered to be $10^{-3} M_P$.}
\label{tab:tabC2}
\end{table}

\vspace{2mm}
\noindent \textbf{Estimate of decay width of inflaton:} Once inflation ends, the inflaton field begins to oscillate around the minimum of its potential and subsequently transfer of inflaton energy density to radiation takes place.
Note that, in Palatini case when the inflaton field is sub-Planckian (or near the minimum) we can approxiamtely write $S\simeq\phi+v_S$ from Eq.(\ref{eq:PalatiniTransO}). In the metric case the field transformation from Jordan to the new frame is a bit nontrivial. We can still write $S\simeq\phi+v_S$ in the limit $\frac{6\xi_S^2v_S^2}{M_P^2}\ll 1$ which becomes $v_S^2\ll \mathcal{O}(10^{31})$ GeV$^2$ for $\xi_S=200$. This bound on $v_S$ gets relaxed for smaller $\xi_S$.

Here we are assuming that the inflaton decays perturbatively into radiation. In the new frame, after the conformal transformation, all the bare mass terms recieve a correction for example $M_{N}^{\rm new}\sim \frac{M_N}{\Omega^2}$.
But as we know the maximum particle production in perturbative decay mode dominantly takes place near the minimum of the inflaton potential \cite{Kofman:1997yn} where $\Omega(S)^2\rightarrow 1$, we can safely use the bare masses of the inflaton and the other fields as originally introduced in the Lagrangian. The conformal factors also can give rise to new interactions of dark matter with inflaton which is heavily suppressed approximately by a factor $
\mathcal{O}\left(\frac{m_\chi^2}{M_P^2}\right)$ in the sub-Planckian inflaton regime.
In our case, the inflaton can decay into SM Higgs pair or to pair of DM. However the latter decay mode can give rise to production of the dark matter non-thermally. Since we focus on purely gravitational production of dark matter, we consider the associated coupling $\lambda_{\chi\phi}$ to be negligible, as mentioned earlier.

We also have right handed neutrinos in our scenario. The mass of the right handed neutrino is generated by a higher dimensional global $U(1)_L$ breaking operator mentioned before. Hence inflaton can also decay into right handed neutrinos if kinematically allowed. We denote the inflaton decay width $\Gamma$ which can be expressed as (when $\phi$ is close to its minimum $v_S$)
\begin{align}
&\Gamma \simeq \Gamma(\phi\rightarrow H H)+ \Gamma(\phi\rightarrow N N)\\
&\Rightarrow \Gamma\simeq\frac{\lambda_{S H}^2 v_S^2}{8\pi m_\phi}\Theta(m_\phi-m_H)+\frac{Y_N^2 v_S^2 m_\phi}{8\pi M_P^2}\Theta(m_\phi-M_N),
\end{align}
In general the $Y_N$ is a $3\times 3$ matrix. For the moment let us consider the right handed neutrino mass matrix to be diagonal with the hierarchy $Y_{N_{33}}\gg Y_{N_{22}},Y_{N_{11}}$. Hence the inflaton decay width is a function of $(\lambda_S, \lambda_{S H},v_S,Y_{N_{33}})$. For making a numerical estimate of $\Gamma$ we first fix $M_N\sim 10^{14}$ GeV, typical type I seesaw scale. With these inputs, we furnish the decay width of the inflaton 
for each of the benchmark points tabulated in tables \ref{tab:tabC1} and \ref{tab:tabC2}. The purpose of the present discussion is to enquire whether a large decay width of the inflaton can indeed be obtained within our proposed setup by varying the model parameters. This discussion ensures that DM production occurs after reheating is complete which is in accordance with the approach we follow here to determine the DM relic density as explained in the upcoming section.

%\begin{figure}[h]
%\noindent \begin{centering}
%\includegraphics[height=7cm,width=8cm]%{Decay1.pdf}~~~~
%\includegraphics[height=7cm,width=8cm]{decay2.pdf}
%\par\end{centering}
%\caption{\label{fig:inflatonDecay} Estimate of decay width of inflaton considering (left panel) $v_S=10^{15}$ GeV and (right panel) $v_S=10^{16}$ GeV. The right handed neutrino masses have been fixed at $10^{14}$ GeV for both the plots.}
%\end{figure}

\vspace{-.2cm}

%%%%%%%%%%%%%%%%%%%%%%%%%%%%%%%%%%%%%%%%%%%
\section{Gravitational Production of Dark matter}
\label{sec3}
%%%%%%%%%%%%%%%%%%%%%%%%%%%%%%%%%%%%%%%%%%%
The setup we have described above opens up the possibility of DM production from purely gravitational effects. We consider $\lambda_{\chi H}$ and $\lambda_{\chi\phi}$ very tiny such that no DM production occurs thermally or non-thermally from decay, annihilations of SM Higgs or inflaton. We describe the dark matter production for both the Palatini and metric formalisms separately as mentioned earlier. We particularly focus on classical production mechanism of dark matter \cite{Laulumaa:2020pqi, Babichev:2020xeg}. In the latter part we shall comment on quantum production and its possible contribution in altering the estimates made from the classical one. For the DM analysis we 
prefer to work with larger $v_S$ (although sub-Planckian) as can be seen from tables \ref{tab:tabC1} and \ref{tab:tabC2} although we find that these choices have very minor impact on the inflationary predictions.
% until it is sub-Planckian (\textcolor{red}{Is $10^{17}$ GeV not sub-Planckian? Or you %mean until inflaton becomes sub-Planckian?}). 
Such choices of $v_S$ are also motivated to obtain a decay width of inflaton larger than $10^{9}$ GeV which ensures that the DM field starts behaving like matter after the reheating of the universe gets over. %\textcolor{red}{As noted in \cite{Laulumaa:2020pqi}, such choices of inflaton decay width also ensures that the amplitude of perturbations equals the observed value at the pivot scale $k=0.05 \; {\rm Mpc}^{-1}$ \cite{Aghanim:2018eyx}.}
In the present work, we solely focus on such kind of scenario.

\subsection{DM production for Palatini formalism} 
The singlet $\chi$ is identified with the dark matter candidate and has the action given by

\begin{equation}
S_{\chi}=\int d^{4}x\sqrt{-g}\bigg(-\frac{g^{\mu\nu}}{2}(\partial_{\mu}\chi)(\partial_{\nu}\chi)-\frac{1}{2}m_{\chi}^{2}\chi^{2}+\frac{\xi_{\chi}}{2}g^{\mu\nu}R_{\mu\nu}\chi^{2}-\frac{\lambda_{\chi}}{4}\chi^{4}\bigg)\label{eq:1}
\end{equation}
After conformal transformation (equation \eqref{conftrans}), its action is redefined as:
\begin{equation}
S_{\chi}^{E}=\int d^{4}x\sqrt{-\hat{g}}\bigg(-\frac{\hat{g}^{\mu\nu}}{2\Omega^{2}\big(S(\phi)\big)}(\partial_{\mu}\chi)(\partial_{\nu}\chi)-\frac{1}{2}\frac{m_{\chi}^{2}\chi^{2}}{\Omega^{4}\big(S(\phi)\big)}+\frac{\xi_{\chi}}{2\Omega^{2}\big(S(\phi)\big)}\hat{g}^{\mu\nu}\hat{R}_{\mu\nu}\chi^{2}-\frac{\lambda_{\chi}}{4\Omega^{4}\big(S(\phi)\big)}\chi^{4}\bigg),
\end{equation}
Note that in Palatini gravity, $R_{\mu\nu}$ doesn't change upon conformal transformation \cite{Karam:2020rpa}. To make the kinetic term of $\chi$ canonical, we make the following redefinition for the $\chi$ field as
\begin{equation}
\sigma=\Omega^{-1} \chi,\label{DMfieldredef}	
\end{equation}
which gives
\begin{align}
S_{\sigma} & =\int d^{4}x\sqrt{-\hat{g}}\bigg(-\frac{\hat{g}^{\mu\nu}}{2}(\partial_{\mu}\sigma)(\partial_{\nu}\sigma)-\frac{\hat{g}^{\mu\nu}\sigma}{\Omega}(\partial_{\mu}\sigma)(\partial_{\nu}\Omega)-\frac{\hat{g}^{\mu\nu}}{2\Omega^{2}}(\partial_{\mu}\Omega)(\partial_{\nu}\Omega)\sigma^{2}-\frac{1}{2}\frac{m_{\chi}^{2}\sigma^{2}}{\Omega^{2}} \nonumber \\
& +\frac{\xi_{\chi}}{2}\hat{R}\sigma^{2}-\frac{\lambda_{\chi}}{4}\sigma^{4}\bigg)
\label{DMaction}
\end{align}
where we have dropped the field dependence for $\Omega$. For $\xi_\chi>0$ and $\frac{{m_\chi}^{2}}{\Omega^2}<12{\xi_\chi}H^{2}$, the effective potential of
$\sigma$ develops a $Z_{2}$ symmetry breaking minimum during inflation 
\begin{equation}
\sigma=\pm\sigma_{*}=\bigg(\frac{12{\xi_\chi} H^{2}-\frac{m_\chi^{2}}{\Omega^2}}{\lambda_\chi}\bigg)^{1/2}\label{eq:2},
\end{equation}
where we use $\hat{R}\simeq 12 H^2$ during inflation, $H$ being the Hubble parameter.

\begin{figure}[h]
\noindent \begin{centering}
\includegraphics[height=6.5cm,width=8cm]{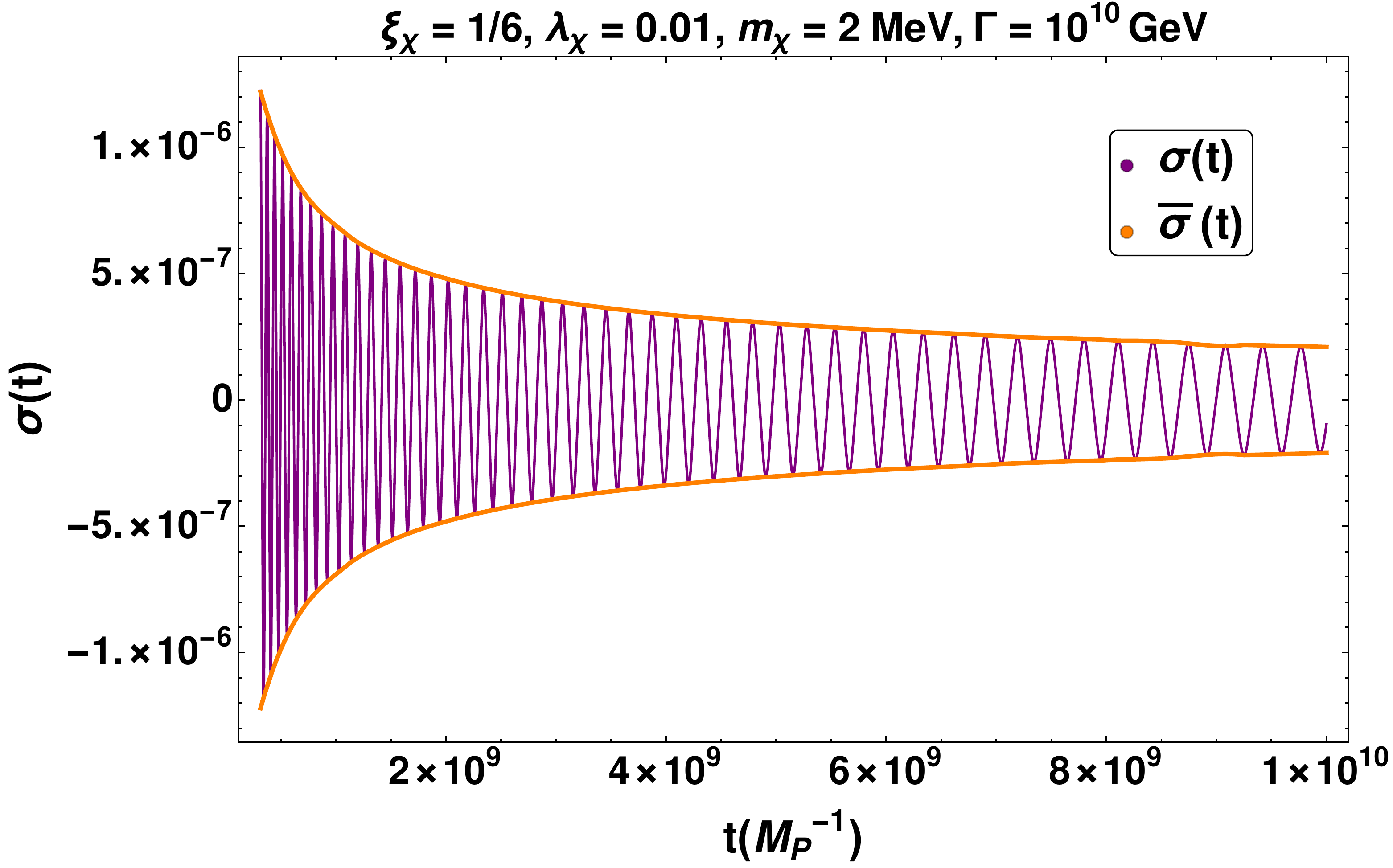}~~~~
\includegraphics[height=6.5cm,width=8cm]{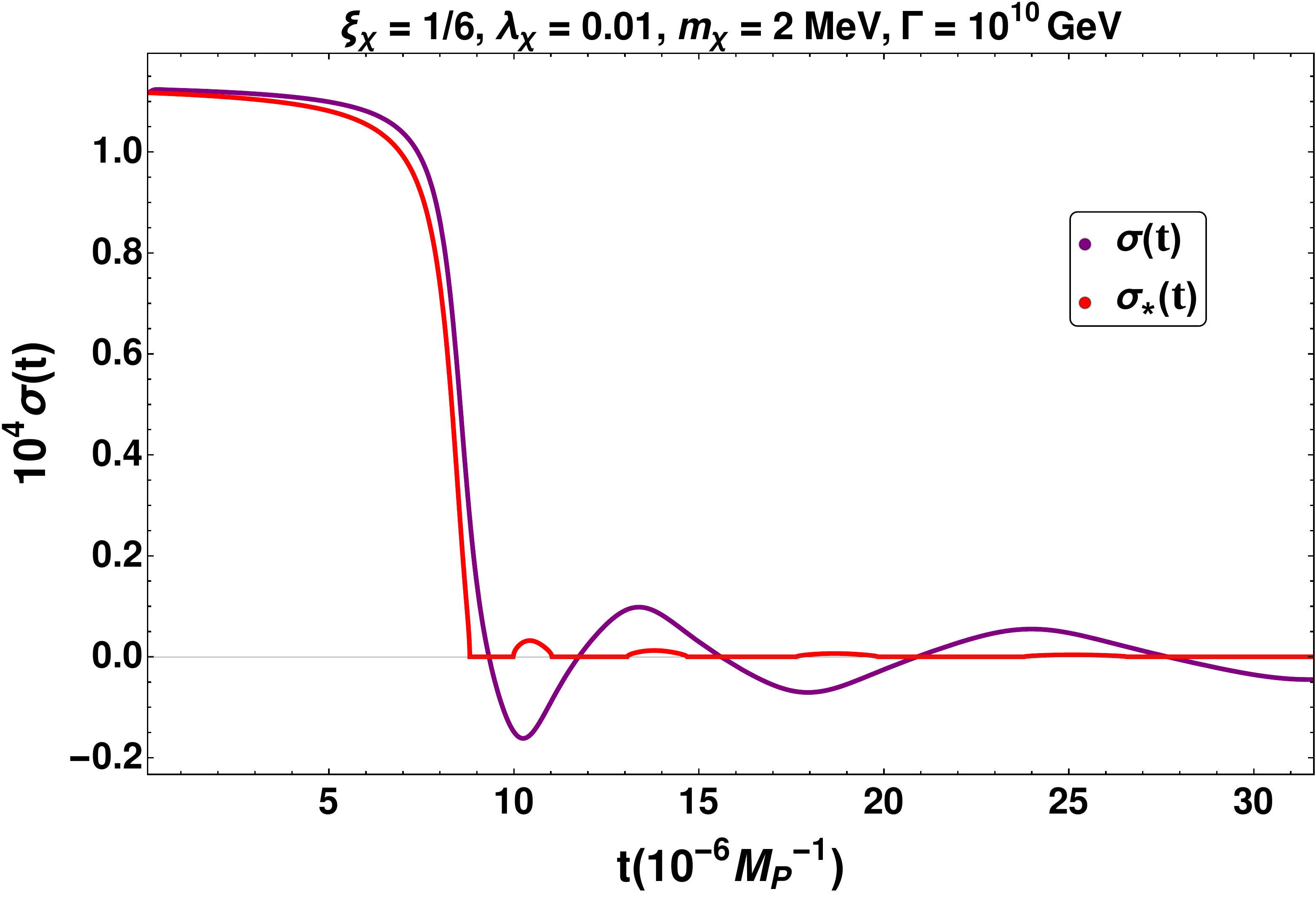}
\par\end{centering}
\caption{\label{fig:Oscillations-of-the-DM-field}(Left panel) Oscillating behaviour of the DM field is shown as function of time. (Right panel) The evolution of curvature induced symmetry breaking  minimum $\sigma_*$ with time is presented. We use the benchmark point CI-3 in table \ref{tab:tabC1} to generate the figures.}
\end{figure}

During inflation, $\sigma$ is located at the curvature induced local minimum $\sigma_{*}$ which is a function of the  Ricci scalar ($\hat{R}(t)$). After inflation, new $Z_2$ symmetry preserving minimum of $\sigma$ appears (with $\langle\sigma\rangle=0$) and the $\sigma$ field starts oscillating (Fig. \ref{fig:Oscillations-of-the-DM-field}) around the minimum with the initial amplitude $\sigma_*$.

Now, after a sufficient number of oscillations, when the potential of $\sigma$ becomes dominated by the quadratic term, we are left with a dark relic component whose energy density scales like non-relativistic matter. This transition epoch is denoted by $t=t_{\rm tr}$. We denote the equality of radiation and matter in the form of $\sigma$ as 
$$ \lambda_\chi\bar{\sigma}_{t_{\rm tr}}^4/4=1/2~ \frac{m_\chi}{\Omega^2}^2 \bar{\sigma}_{t_{\rm tr}}^2 $$ where $\bar{\sigma}$ denotes the envelope of the oscillating $\sigma$ field. Utilising this condition and the fact that conformal factor $\Omega=1$ at such late times, we can find $\bar{\sigma}$ at the transition point as: $\bar{\sigma}_{t_{\rm tr}}=\left(\frac{2 m_\chi^2}{\lambda_\chi}\right)^{1/2}$. Here it is worth mentioning that we will work with $t_{\rm tr}\gg t_{\rm reh}$ case, where $t_{\rm reh}$ is the time at the end of reheating \footnote{In principle $t_{\rm tr}$ could be bigger or smaller than the reheating time of the universe, however in the later case the evolution history of the DM would be different from the present one, as discussed in \cite{Babichev:2020xeg} for superheavy dark matter.}. Satisfying this inequality would depend on the inflaton decay width, quartic coupling $\lambda_\chi$ and mass of the DM.
Subsequently, with the $\sigma$ field behaving like matter, the envelope of its oscillation at present time $t\gg t_{\rm tr}$ can be obtained using following equalities :
\begin{align}
&\overline{\sigma(t)} a(t)^{3/2}=\bar{\sigma}_{t_{\rm tr}} a({t_{\rm tr}})^{3/2},\label{DMenv1}\\
& a({t_{\rm tr}})\bar{\sigma}_{t_{\rm tr}}=a(t_{\rm reh}) \bar{\sigma}_{\rm reh}.
\label{DMenv2}
\end{align}
We find,
\begin{equation}
\overline{\sigma(t)}=\left(\frac{\lambda_{\chi}}{2}\right)^{1/4}\frac{\overline{\sigma_{{\rm reh}}}^{3/2}}{m_{\chi}^{1/2}}\left(\frac{a_{{\rm reh}}}{a(t)}\right)^{3/2}\label{sigmapresent}
\end{equation}
With this, the present relic density of matter like $\sigma$ field is obtained as 
\begin{align}
  &\rho_\sigma(t)=\frac{1}{2} m_\chi^2\overline{\sigma(t)}^2 \\
\Rightarrow &~\Omega_{\sigma}(t)h^{2}=\frac{\sqrt{\lambda_\chi}}{2\sqrt{2}}\frac{m_\chi\overline{\sigma_{1}}^{3}}{3(H_{0}/h)^{2}M_{P}^{2}}\frac{g_{*s,0}T_{0}^{3}}{g_{*s,1}T_{1}^{3}},\label{eq:3}
\end{align}
where suffix ``1" in temperature denotes a time $t_1$ in radiation dominated universe which is close to the end of reheating phase and much lesser than $t_{\rm tr}$, so that we can safely use the above relations (equation \eqref{DMenv1}-\eqref{sigmapresent}). The $\overline{\sigma_{1}}$ indicates the envelope of the oscillating DM field at $t=t_{1}.$ We make a conservative choice of $t_{1}$ where $H(t_{1})\cong0.01\Gamma$ and $\Gamma$ is the decay width of the inflaton.

The relevant set of coupled Boltzmann equations (BEs) involving different fields is given by
\begin{align}
&\ddot{\phi}+3H\dot{\phi}+\Gamma\dot{\phi}+V'(\phi) =0 \\
&\ddot{\sigma}+3H\dot{\sigma}+\lambda_\chi\sigma^{3}+\frac{m_\chi^{2}}{\Omega^2}\sigma-\xi_\chi \hat{R}\sigma+\frac{3H\sigma\dot{\Omega}}{\Omega}+\frac{\sigma\ddot{\Omega}}{\Omega}-\frac{2\sigma(\dot{\Omega})^{2}}{\Omega^{2}}  =0 \label{DMeqn}\\
&\dot{\rho_{r}}+4H\rho_{r}  =\Gamma\dot{\phi}^{2} \\~~~~~~~
%(\text{\color{blue} Confusion II:}
 %\text{\color{red}Does $\Omega$ enter here?}) \\
&3H^{2}M_{P}^{2} =\frac{1}{2}\dot{\phi}^{2}+V(\phi)+\rho_{r}\\
&\hat{R}  =M_{P}^{-2}\bigg(4V(\phi)-\dot{\phi}^{2}\bigg). 
\end{align}
For the inflaton field, we have considered the slow roll initial conditions  and for
the DM candidate $\sigma,$ we take $\sigma(t_{\rm in})=\sigma_{*}(t_{\rm in})$ and $\dot{\sigma}(t_{\rm in})=0$ as the initial conditions. Note that, the BE of the DM field includes the conformal factor $\Omega$, which was not considered in earlier works. Also after the end of inflation when the inflaton field turns sub-Planckian, the role of $v_{S}$ cannot be ignored. Therefore, to determine the Hubble parameter and evolution of Ricci scalar in the post-inflationary era one should work with the exact potential in the new frame which resembles the following form:
\begin{align}
 V_N(S(\phi))\simeq \frac{\frac{\lambda_S(\phi)}{4}\left[S(\phi)^2-v_S^2\right]^2}{\left[1+\frac{\xi_S \{S(\phi)^2-v_S^2\}}{M_P^2}\right]^2}+\frac{\kappa \left[S(\phi)^2-v_S^2\right]^2{\rm Log}\left[\frac{S(\phi)}{\mu}\right]}{\left[1+\frac{\xi_S \{S(\phi)^2-v_S^2\}}{M_P^2}\right]^2},
 \label{eq:oriP}
\end{align}
with $S(\phi)$ given by Eq.(\ref{eq:PalatiniTransO}) for the Palatini case.
This clearly tells that the Hubble parameter and subsequently the Ricci scalar have explicit dependence on the inflationary non-minimal parameter $\xi_S$.

\begin{figure}[h]
\includegraphics[height=7cm,width=8cm]{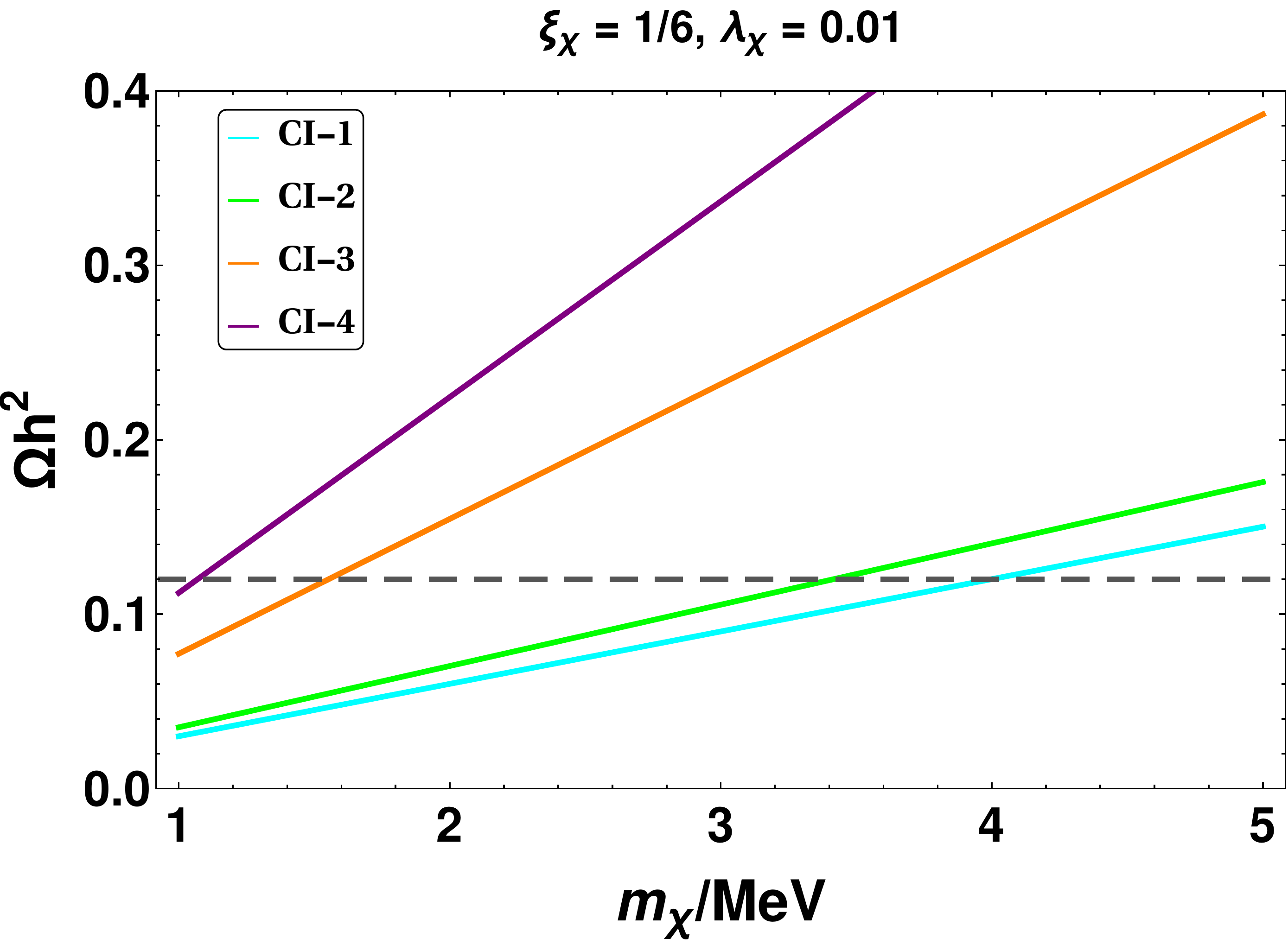}~~~
\caption{ Relic density versus DM mass for $\xi_\chi=\frac{1}{6}$ considering only the classical production for four benchmark points shown in table \ref{tab:tabC1}. The dashed line represents the observed relic abundance in the universe.}
\label{mvsrelicP}
\end{figure}

\begin{figure}[h]
\includegraphics[scale=0.3]{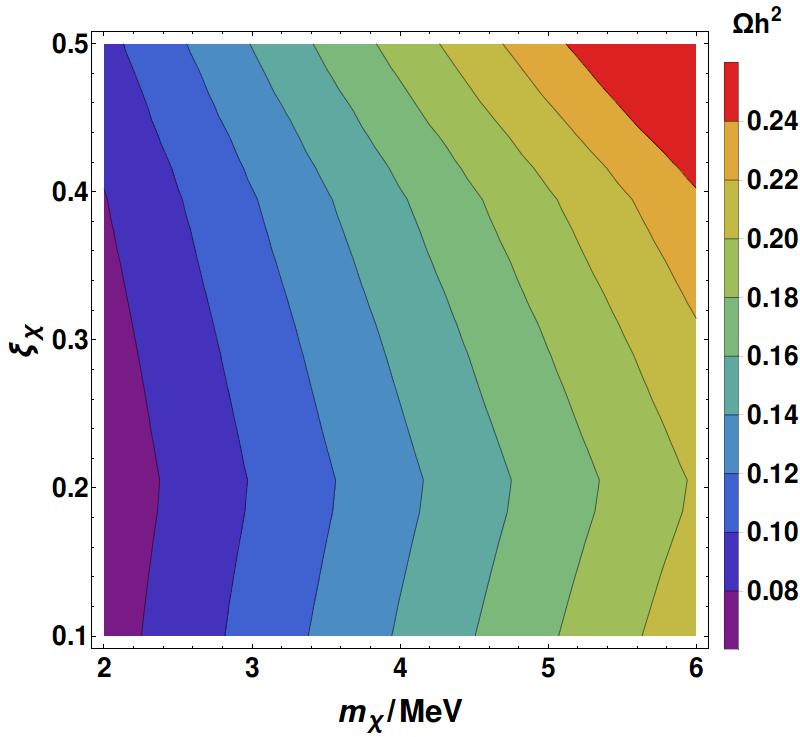}~~~
\includegraphics[scale=0.3]{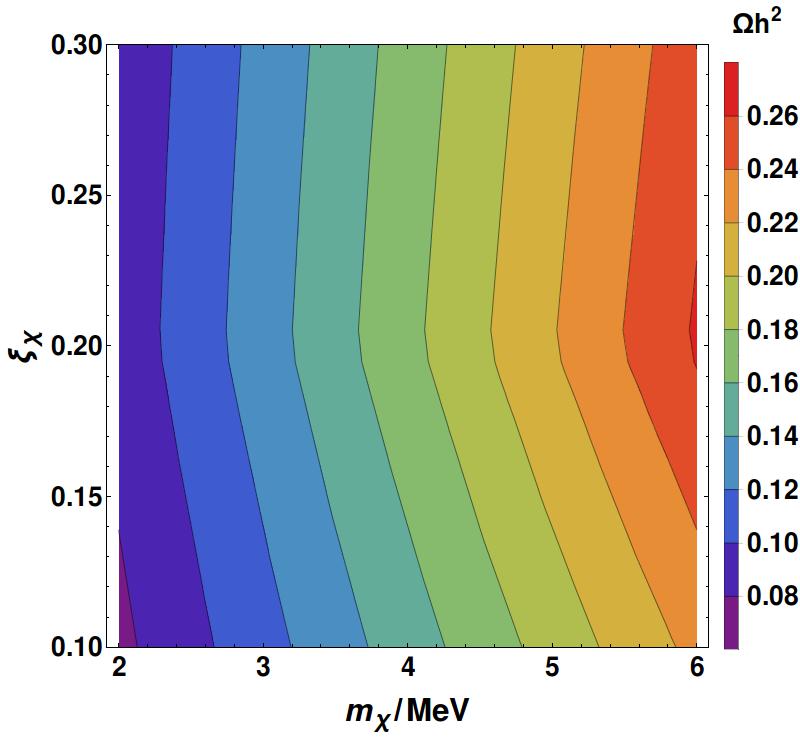}
\caption{Relic density contour lines in $\xi_\chi-m_\chi$ plane for benchmark points CI-3 (left panel) and CI-1 (right panel).}
\label{contourplotP}
\end{figure}

Solving the above coupled Boltzmann equations upto the present epoch becomes too heavy numerically and hence while performing the numerical analysis, we 
solve the above equations upto the time $t_1$ and find the present relic abundance using equation \eqref{eq:3}. We display the evolution of the oscillating DM field as function of time in left panel of Fig. \ref{fig:Oscillations-of-the-DM-field} for the benchmark point CI-3 in table \ref{tab:tabC1}. The oscillation shows a natural damping pattern where the amplitude of oscillation decreases with time. The right panel provides the evolution of symmetry breaking minimum $\sigma_*$ as function of time. It is observed that as time increases $\sigma_*$ tends to reach zero, {\it i.e.}, its global minimum.

In Fig. \ref{mvsrelicP}, we present the relic density variation as function of DM mass for all four benchmark points presented in table \ref{tab:tabC1} by considering $\xi_\chi=1/6$ which corresponds to the conformal limit. The figure exhibits that the DM relic gets enhanced with DM mass, as expected from equation \eqref{eq:3}. It is also observed that the relic abundance of DM depends on both non-minimal inflationary parameters as well the decay width of inflaton.
In general, we confer that 
with $\lambda_\chi=0.01$, $\xi_\chi\sim 1.6$, $\Gamma\sim 10^{10}-10^{11}$ GeV, we obtain the correct relic for DM mass $\mathcal{O}(1)$ MeV.

%We find that increasing $\xi_S$ leads to enhanced DM relic. 

%Therefore, DM relic does depend upon parameters in the inflationary sector and also on the type of inflationary model which, in this part of our analysis, is considered to be quartic inflation with non-minimal coupling to gravity.

In the above analysis for Case I, we have kept the $\xi_\chi$ fixed at the conformal value 1/6.  We extend our analysis for the non-conformal case {\it i.e.} $\xi_\chi\neq \frac{1}{6}$ in Fig. \ref{contourplotP}. We use the benchmark points CI-I and CI-3 from table \ref{tab:tabC1} for the purpose. The dependence of the relic abundance on $\xi_\chi$ does not show a regular pattern as also noted in \cite{Laulumaa:2020pqi}. Apparently, from equation \eqref{eq:2} it seems that a larger $\xi_\chi$ should increase the initial amplitude of $\sigma$ oscillation. Now the presence of the non-minimal ${\xi_\chi}\hat{R}\sigma^2$ term in the Lagrangian provides a dynamical mass contribution to the DM. Initially, a large $\xi_\chi$ with non-vanishing $R$ makes the $\chi$ field more massive and hence it tracks the non-trivial minimum $\sigma_*$ for a longer time, eventually leading to a smaller initial amplitude. Also, the oscillating behaviour of $\hat{R}$ during reheating may slow down or accelerate the red-shifting of the amplitude of $\sigma$ oscillation, depending on the phase difference between $\hat{R}(t)$ and $\sigma(t)$. Hence, due to combination of these three effects, we find irregular behaviours of relic contours while we enhance the $\xi_\chi$ for a fixed $m_\chi$, with the choice $\lambda_\chi\sim 0.01$. For example in the left panel of Fig. \ref{contourplotP}, when $0.2\lesssim\xi_\chi\lesssim0.5$, we observe that with the increase of $\xi_\chi$ it requires smaller DM mass to obey the observed relic limit. Below $\xi_\chi<0.2$, the pattern however gets reversed. Similar irregular pattern is observed for a comparatively larger $\Gamma$ as well in the right panel of Fig. \ref{contourplotP}. %{\color{blue}Note that the effect of increasing $\Gamma$ is to decrease the  number of oscillations needed to complete reheating, leading to a higher reheating temperature. Thus, the effect of the oscillating $\hat{R}(t)$ term lasts for a shorter time. This can, in general, either increase or decrease the relic; although for our choice of parameters we see that the relic increases.{\color{red}(CHECK?).}} 

Here it is important to mention that the condition $V''(\sigma_*)>9/4~H^2$ is required to be maintained from the non-observation of isocurvature perturbations \cite{Akrami:2018odb}. This translates into a lower bound on $\xi_\chi$ ($\gtrsim 0.1$) \cite{Laulumaa:2020pqi}. Another pertinent point is regarding the lower bound on the DM mass which arises from the condition that the temperature during radiation to matter transition behaviour for the DM field should be larger than the matter-radiation equality temperature of the universe ($T_{\rm eq}$). To be precise,
\begin{align}
&T_{\rm tr}>T_{\rm eq}\sim 0.75 {~\rm eV}\\
\Rightarrow~ & m_\chi>0.75 {~\rm eV}\times \left(\frac{g_{*}({\rm tr})}{g_*(t_1)}\right)^{1/3}\times \frac{\bar{\chi}(t_1)}{(2/\lambda_\chi)^{1/2}T(t_1)}\label{DMmasslb}
\end{align}
Numerically we have found $m_\chi\gtrsim \mathcal{O}(10^{-3})$ eV for any arbitrary choice of dark sector parameters. 

\begin{figure}[h]
\includegraphics[scale=0.45]{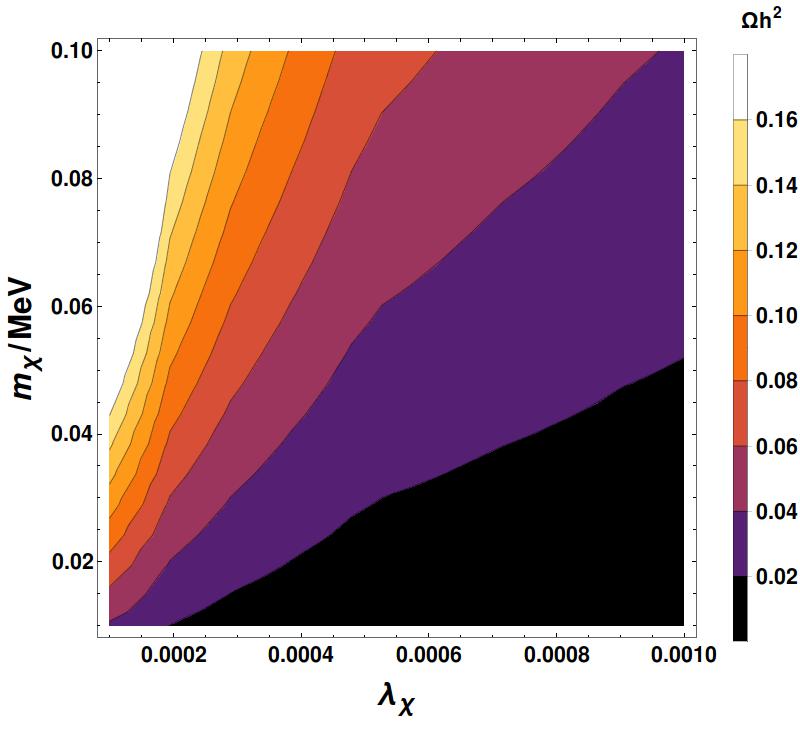}
\caption{Relic density contour lines in $m_\chi-\lambda_\chi$ plane with $\xi_\chi=\frac{1}{6}$. We have utilised the benchmark point CI-3 to generate this figure.}
\label{contourplotlm}
\end{figure}

So far in our analysis we have kept $\lambda_\chi$ fixed at 0.01. A lower bound on $\lambda_\chi$ can be found by demanding that $\sigma_*$ defined in equation \eqref{eq:2} remains sub-Planckian ($\lesssim 0.1 M_P$). This means,
\begin{align}
\lambda_\chi\gtrsim 1.2\times 10^3 ~ \frac{\xi_\chi H_{\rm inf}^2}{M_P^2},\label{eqn:lambdabound}
\end{align}
 where $H_{\rm inf}$ is the Hubble parameter during inflation. Considering $H_{\rm inf}\sim 10^{-5} M_P$, it turns out that $\lambda_\chi\gtrsim 10^{-7}$ with $\xi_\chi\sim \mathcal{O}(1)$. In Fig. \ref{contourplotlm}, we have presented the relic contour lines in $m_\chi-\lambda_\chi$ plane for a fixed $\xi_\chi=1/6$ using the benchmark point CI-3 from table \ref{tab:tabC1}. It is observed that for a certain DM mass, the relic density decreases with the enhancement of $\lambda_\chi$. This happens since the amplitude of $\chi$ oscillation gets reduced when $\lambda_\chi$ rises.
 \begin{figure}[h]
\includegraphics[scale=0.45]{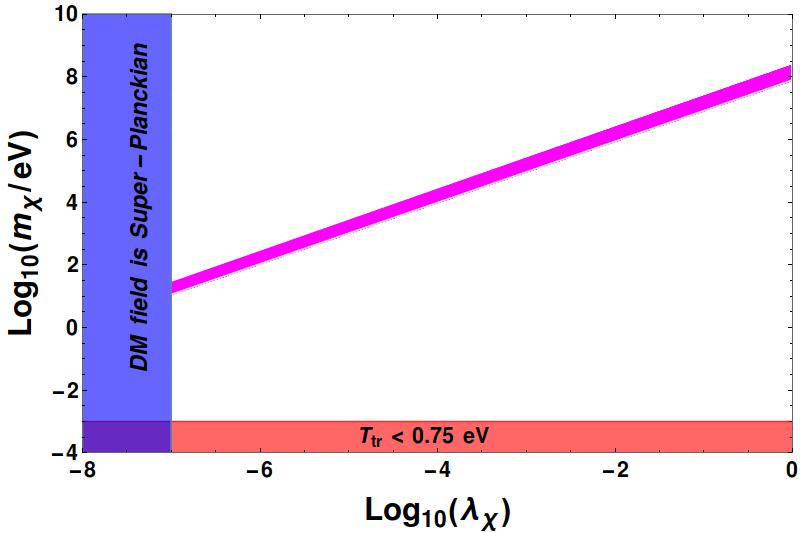}
\caption{DM relic satisfied region is presented in the same plane with $\xi_\chi$  varied in the range of 0.1-1.0 for the benchmark point CI-1 as recorded in table \ref{tab:tabC1}.}
\label{contourplotlm2}
\end{figure}

In Fig. \ref{contourplotlm2} we have considered the variation of $\xi_\chi$ in the range 0.1-1.0 and obtained the relic satisfied parameter space ($\Omega_{\rm DM} h^2=0.12$, magenta shaded) in $m_\chi-\lambda_\chi$ plane. We have utilised the benchmark point CI-1 for the purpose. In the blue coloured region, violation of the inequality as presented in equation \eqref{DMmasslb} occurs. In the red coloured region, the initial amplitude of the DM field value turns super-Planckian (as can be seen from equation \eqref{eqn:lambdabound}).

%\begin{figure}[h]
%\includegraphics[scale=0.44]{contourplotlm}~~~
%\includegraphics[scale=0.27]{contourplotinf}~~~
%\caption{Case I: Relic density contour lines in $m_\chi-\xi_S$ plane with $%\xi_\chi=\frac{1}{6}$, $\lambda_\chi=0.01$ and  $\Gamma=10^{10}$ GeV.}
%\label{contourplotXiS}
%\end{figure}

%%%%%%%%%%%%%%%%%%%%%%%%%%%%%%%%%%%%%%%%%%%%%%
\subsection{DM production for metric formalism}
%%%%%%%%%%%%%%%%%%%%%%%%%%%%%%%%%%%%%%%%%%%%%%%
Next we consider the metric formalism as discussed in section \ref{sec2}. Here, under the conformal transformation (equation \eqref{conftrans}), the Ricci scalar $R$ in the Jordan frame transforms as \cite{Kaiser:2010ps}
\begin{equation}
\hat{R}=\frac{1}{\Omega^{2}}\bigg(R-\frac{6}{\Omega}\frac{1}{\sqrt{-g}}\partial_{\mu}\big(\sqrt{-g}g^{\mu\nu}\partial_{\nu}\Omega\big)\bigg),\label{Rtrans}
\end{equation} 
where $\hat{R}$ is the Ricci scalar at Einstein frame.
Using the above equation and equation \eqref{conftrans}, the action in the Einstein frame after the field redefinitions (equation \eqref{DMfieldredef}) becomes
\begin{align}
S_{\sigma}=&\int d^{4}x\sqrt{-\hat{g}}\bigg[-\frac{\hat{g}^{\mu\nu}}{2}(\partial_{\mu}\sigma)(\partial_{\nu}\sigma)-\frac{\hat{g}^{\mu\nu}\sigma}{\Omega}(\partial_{\mu}\sigma)(\partial_{\nu}\Omega)-\frac{\hat{g}^{\mu\nu}}{2\Omega^{2}}(\partial_{\mu}\Omega)(\partial_{\nu}\Omega)\sigma^{2}-\frac{1}{2}\frac{m_{\chi}^{2}\sigma^{2}}{\Omega^{2}}\nonumber \\
&+\frac{\xi_{\chi}}{2}\hat{R}\sigma^{2}-\frac{\lambda_{\chi}}{4}\sigma^{4}+3\xi_{\chi}\sigma^{2}\bigg(\frac{2\dot{\Omega}^{2}}{\Omega^{2}}-\frac{\ddot{\Omega}}{\Omega}-\frac{3\dot{\Omega}H}{\Omega}\bigg)\bigg] 
\end{align}
Note that the last three terms do not appear in the case I (equation \eqref{DMaction}) discussed earlier. The $Z_2$ symmetry breaking minimum (equation \eqref{eq:2}) now takes the form:
\begin{equation}
\sigma=\pm\sigma_{*}=\bigg(\frac{12{\xi_\chi} H^{2}+12\xi_\chi\frac{\dot{\Omega}^2}{\Omega^2}-\frac{m_\chi^{2}}{\Omega^2}-6\xi_\chi\frac{\ddot{\Omega}}{\Omega}-18\xi_{\chi}H\frac{\dot{\Omega}}{\Omega}}{\lambda_\chi}\bigg)^{1/2}.
\label{eqn:DMminS}
\end{equation}

\begin{figure}[h]
\includegraphics[height=7cm,width=8cm]{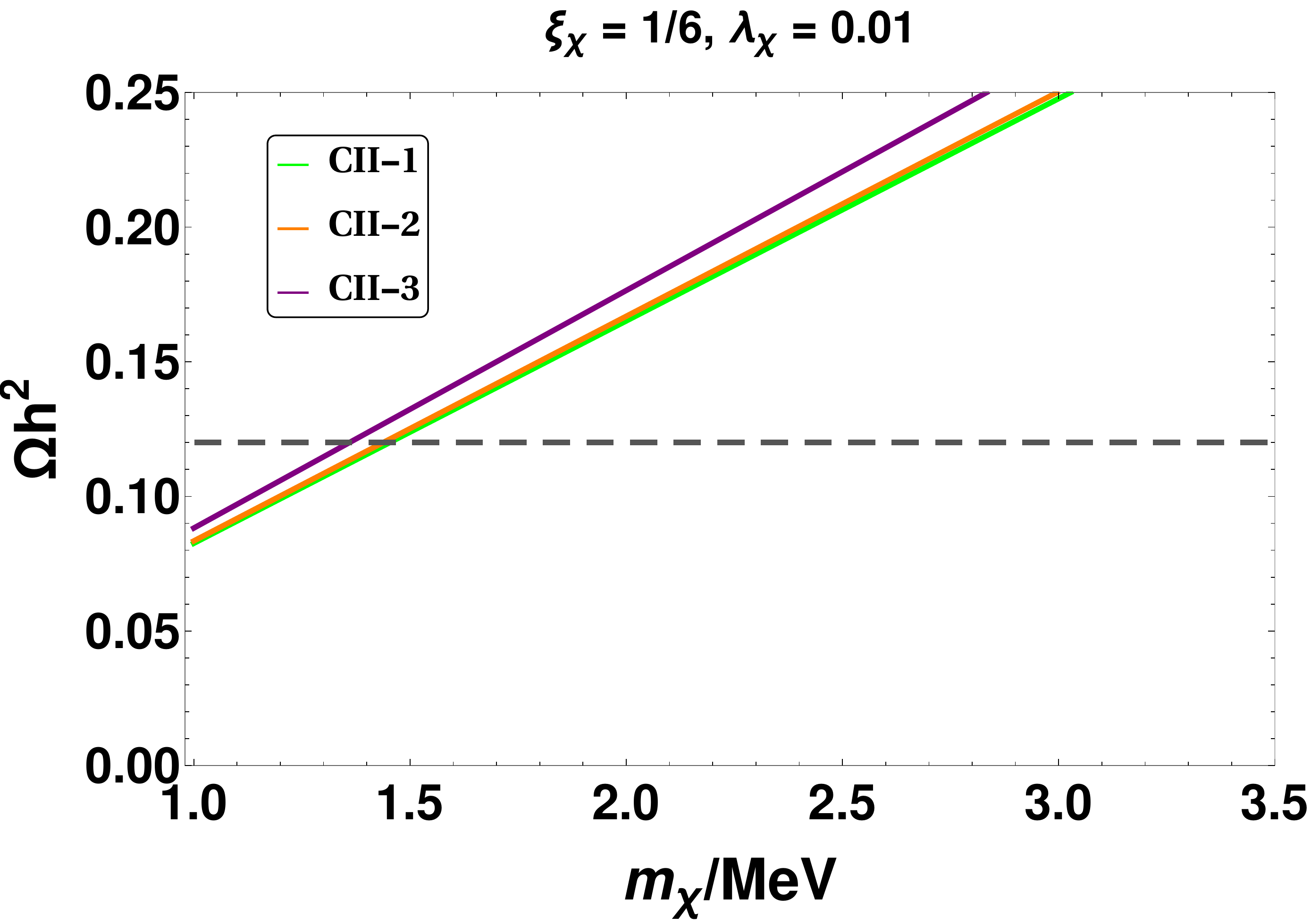}~~~
\caption{Relic density vs DM mass for $\xi_\chi=\frac{1}{6}$ considering only the classical production for three benchmark points listed in table \ref{tab:tabC2}. The dashed band represents the observed relic abundance in the universe.}
\label{mvsrelicS}
\end{figure}

\begin{figure}[h]
\includegraphics[scale=0.35]{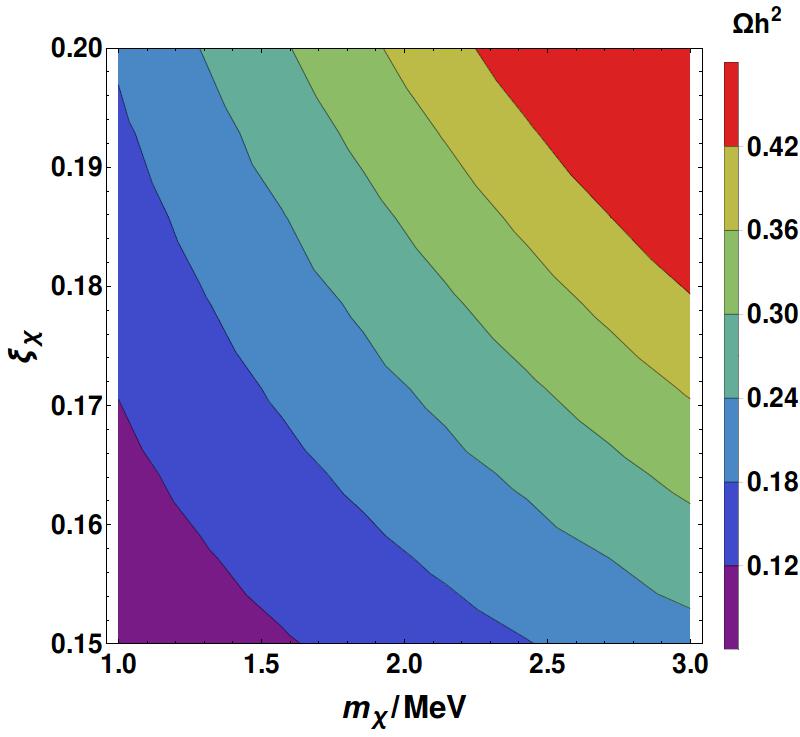}
\caption{Relic density contour lines in $\xi_\chi-m_\chi$ plane by fixing $\lambda_\chi=0.01$ for the reference point CII-3 in table \ref{tab:tabC2}.}
\label{contourplotS}
\end{figure}
\vspace{2mm}
Following the same intermediate steps as in the case I, we can find the present relic abundance of DM using equation \eqref{eq:3}. However, the DM equation of motion (equation \eqref{DMeqn}) gets modified to
\begin{align}
&\ddot{\sigma}+3H\dot{\sigma}+\lambda_\chi\sigma^{3}+\frac{m_\chi^{2}}{\Omega^2}\sigma-\xi_\chi \hat{R}\sigma+\frac{3H\sigma\dot{\Omega}}{\Omega}+\frac{\sigma\ddot{\Omega}}{\Omega}\nonumber\\
&-\frac{2\sigma(\dot{\Omega})^{2}}{\Omega^{2}}+6\xi_{\chi}\frac{\sigma\ddot{\Omega}}{\Omega}+18\xi_{\chi}H\frac{\sigma\dot{\Omega}}{\Omega}-12\xi_{\chi}\frac{\sigma\dot{\Omega}^{2}}{\Omega^{2}}  =0
\label{eqn:DMeomS}
\end{align}

Note that when $\kappa,v_S=0$, as in equation \eqref{Starpot}, the parameters $\xi_S$ and $\lambda_S$ appear only as a ratio in the inflationary potential, which is fixed by the observed value of scalar perturbation spectrum $P_S$. Thus the Hubble paramter parameter  remains unaltered upon changing either of the $(\xi_S,\lambda_S)$. The only effect of $\xi_S$  in DM production arises through the conformal factor appearing in the initial oscillation amplitude of DM (\textit{i.e.} $\sigma_*$) and the set of Boltzmann equations. Therefore, it is expected to obtain some differences in the DM phenomenology compared to analysis with the minimally coupled Starobinsky potential as was the case in \cite{Laulumaa:2020pqi}. In our case (table \ref{tab:tabC2}), in addition to the non-minimal coupling $\xi_S$, we also have non-zero $\kappa$ and $v_S$ which enter into the Hubble parameter. While solving the Boltzmann equations we will use the original potential as defined in equation (\ref{eq:oriP}). The field transformation $S(\phi)$ is non-trivial to obtain analytically from equation (\ref{fieldtrans}) since $f=1$ in the metric case. Hence we follow a numerical technique. %by employing the interpolation method.}  

%the parameters $\xi_S$ and $\lambda_S$ appear only as a ratio, which is fixed by the observed value of scalar perturbation spectrum $P_S$.
%(equation \eqref{Starpot}),  
%Thus, apparently the DM analysis should not depend on the inflationary parameters, particularly $\xi_S$, as was the case in \cite{Laulumaa:2020pqi}. However in presence of non zero $\kappa$ and large $v_S$ 
%we expect some dependence of DM relic density on $\xi_S$.

The relic density variation with mass is shown in Fig. \ref{mvsrelicS} for $\xi_\chi=1/6$ for the benchmark points given in table \ref{tab:tabC2}. It is found that for benchmark points CII-1 and CII-2, the DM relic is satisfied at $m_\chi\sim 1.5$ MeV. Note that this is in sharp contrast with earlier results obtained by the authors\footnote{We could find the similar results as in \cite{Laulumaa:2020pqi}, by making the conformal factor $\Omega\rightarrow 1$ in the initial oscillation amplitude of the dark matter as well as the governing Boltzmann equations.} of \cite{Laulumaa:2020pqi} where with same order of $\xi_\chi$ and $\lambda_\chi$, one requires a DM mass around 10 times larger than what we found to satisfy the observed relic abundance. We find that the presence of conformal factor helps in enhancing the amplitude of $\sigma$ oscillation which in turn, through equation \eqref{eq:3}, increases the DM abundance compared to the case of $\Omega \sim 1$ (the case shown in ref. \cite{Laulumaa:2020pqi}). 
 
\begin{figure}[h]
\includegraphics[scale=0.45]{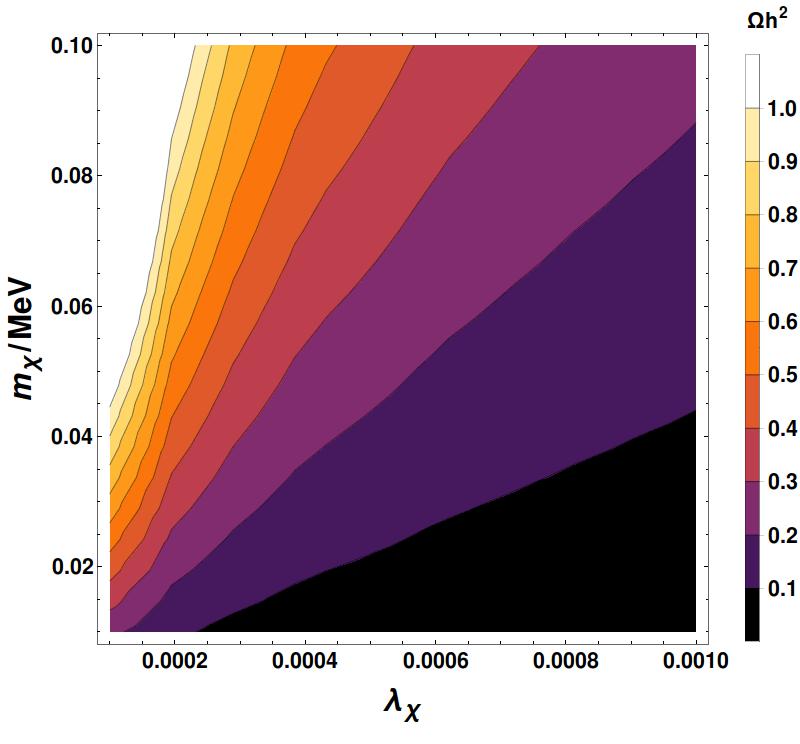}~~
\caption{Relic density contour lines in $\lambda_\chi-m_\chi$ plane with fixed values of $\xi_\chi=1/6$ for the reference point CII-3 in table \ref{tab:tabC2}.}
\label{fig:contourplotSlm}
\end{figure} 

\begin{figure}[h]
\includegraphics[scale=0.45]{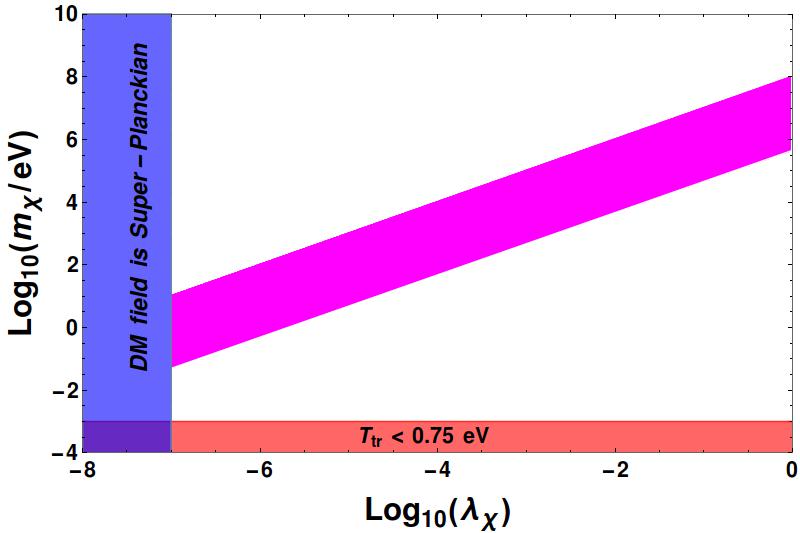}
\caption{Relic satisfied region is presented in t$\lambda_\chi-m_\chi$ plane with $\xi_\chi$  varied in the range of 0.1-1.0 for the reference point CII-3 in table \ref{tab:tabC2}.}
\label{fig:contourplotSlm2}
\end{figure} 
 
Hereafter we work with the benchmark point CII-3. In Fig. \ref{contourplotS}, we show different DM abundance contour lines in the $m_\chi-\xi_\chi$ plane considering $\lambda_\chi=0.01$. In the range of $\xi_\chi$ shown in the plot, we find that for a fixed DM mass, with the enhancement of $\xi_\chi$, DM abundance increases. Keeping $\xi_\chi$ fixed at 1/6, we also show different relic contours in $\lambda_\chi-m_\chi$ plane in Fig. \ref{fig:contourplotSlm}, which exhibits a similar behaviour as found for Case I.  In Fig. \ref{fig:contourplotSlm2} we have considered the variation of $\xi_\chi$ in the range 0.1-1.0 as in Palatini formalism and obtained the relic satisfied parameter space ($\Omega_{\rm DM} h^2=0.12$, magenta shaded) in $m_\chi-\lambda_\chi$ plane. The blue coloured region violates the inequality as presented in equation \eqref{DMmasslb}. In the red coloured region, the initial amplitude of the DM field value turns super-Planckian (as seen from equation \eqref{eqn:lambdabound}).

%{Finally, we study the dependence of the DM relic on the inflationary parameter $\xi_S$. Note that in the Starobinsky potential (eqn. \ref{Starpot}), the parameters $\xi_S$ and $\lambda_S$ appear only as a ratio, which is fixed by the observed value of scalar perturbation spectrum $P_S$. Thus, the DM analysis should not depend on the inflationary parameters, as was the case in \cite{Laulumaa:2020pqi} unless $v_S$ is very large. In large $v_S$ limit the potential takes the form 
%\begin{equation}
%V(\phi)=\frac{\frac{\lambda_{S}}{4}\bigg(\big(\exp(\sqrt{\frac{2}{3}}\frac{\phi}{M_{P}})-1\big)\frac{M_{P}^{2}}{\xi_{S}}-v_{S}^{2}\bigg)^{2}}{\big(\exp(\sqrt{\frac{2}{3}}\frac{\phi}{M_{P}}\big)}\label{StarPotwithvs}
%\end{equation}  
%Recall that, $v_S$ has negligible role in inflationary dynamics however this leads to non-trivial dependence of DM relic abundnace on the inflationary parameter $\xi_S$. We show the relic contours in Fig. \ref{fig:contourplotXiS2} considering $v_S\sim 10^{17}$ GeV and correspondingly, $\Gamma=10^{12}$ GeV.  }

%\begin{figure}[h]
%\includegraphics[scale=0.4]{contourplotlmS.jpeg}
%\includegraphics[scale=0.27]{contourplotinfS.jpeg}
%\caption{Case II: Relic density contour lines in $\xi_S-m_\chi$ plane with fixed values of $\xi_\chi=1/6$, $\lambda_\chi=0.01$ and $\Gamma=10^{12}$ GeV.}
%\label{fig:contourplotXiS2}
%\end{figure}
%\vspace{2mm}

Till now, our focus has remained on the classical production of dark matter.  The classical gravitational production of DM takes place through the misalignment mechanism owing to the presence of quartic term in the Lagrangian for the DM. However one may have gravitational quantum production of DM \cite{Chung:1998ua, Chung:1998zb, Kuzmin:1998uv, Kuzmin:1998kk, Greene:1997ge, Chung:1998rq, Garny:2015sjg, Ahmed:2020fhc, Kolb:2020fwh,Ema:2018ucl,Li:2019ves,Velazquez:2019mpj,Li:2020xwr,Haro:2019umj,Lankinen:2019ifa,Ema:2019yrd,Haro:2019ndy,deHaro:2019oki,Haro:2018zdb} also in the present scenario. Hence, in principle it is expected to take into account both the classical and quantum contributions while computing the DM  relic. Earlier it was found that the DM needs to be superheavy ($m_{\rm DM}\gtrsim H_{\rm inf}$) to yield the correct relic abundance if generated solely through quantum process \cite{Chung:1998zb,Babichev:2020yeo}. That also means that in the low  DM mass range, the quantum production gives rise to under-abundance. On the other hand, we have found that for $m_{\rm DM}\sim \mathcal{O}(1)$ MeV, classical production of DM can provide correct order of relic abundance. We have also shown that the relic increases linearly with the DM mass. Therefore it is safe to state that classical production dominates over the quantum one within our working range of associated parameters including low DM mass. 

It is also worth mentioning that with larger $\xi_\chi(\gg 1)$, the frequency of the Fourier modes for DM oscillation could turn negative. Then, explosive DM production can occur in the tachyonic direction \cite{Fairbairn:2018bsw}. To avoid that, we have worked with $\xi_\chi<1$ in the present framework.

\section{Extension to other inflationary models}
\label{sec4}
Here we briefly discuss the classical gravitational production of DM considering two different inflation models namely, natural inflation and hilltop inflation with the inflaton minimally coupled to gravity. The purpose of this is to check if the above conclusion changes for a different inflationary model, still allowed by Planck 2018 data \cite{Akrami:2018odb}.

First, we assess the case for natural inflation \cite{Freese:1990rb,Adams:1992bn}. Since the natural inflation in its minimal form is ruled out by Planck 2018 data, authors of ref. \cite{Nomura:2017ehb} proposed a modified form of it with the potential given by
\begin{align}
V(\phi)=M^4\Bigg[1-\frac{1}{\left\{1+ \left(\frac{\phi}{M_P}\right)^2\right\}^p}\Bigg],~~~~~~~p>0.
\end{align} 
This modified version of natural inflation predicts values of $n_s$ and $r$ which are allowed by Planck 2018 data for certain range of parameter space. For example, with $p=3,~ M\sim 1.8\times 10^{-3}M_P$ the spectral index and tensor to scalar ratio come out to be 0.97 and $3\times 10^{-4}$ respectively.

\begin{figure}
\includegraphics[height=7cm,width=8cm]{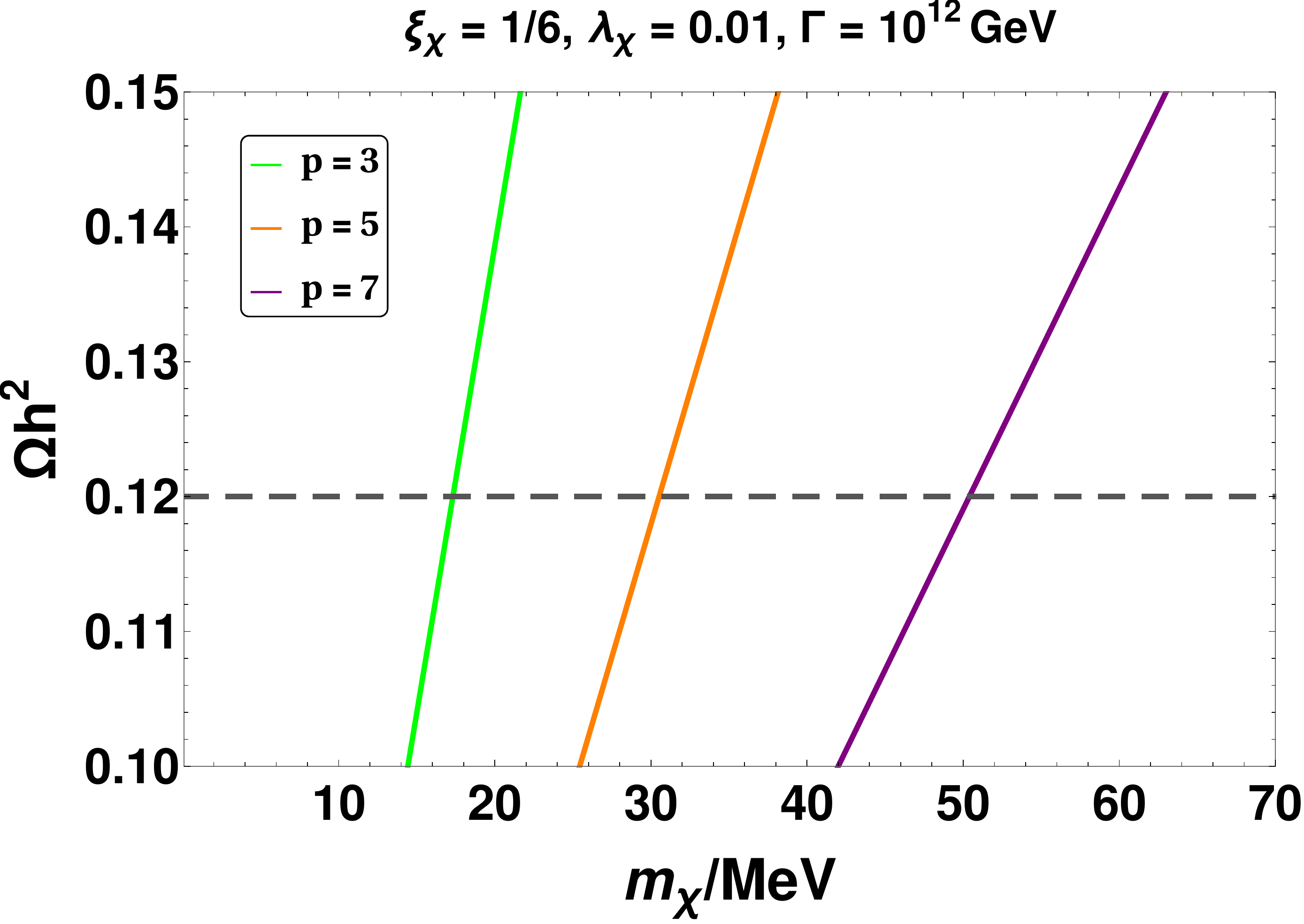}~~
\includegraphics[height=6.7cm,width=8cm]{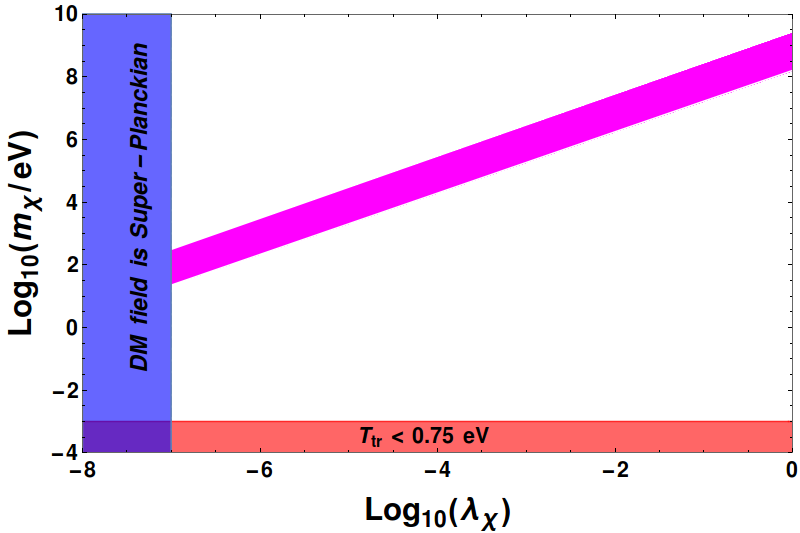}
\caption{(Left panel) DM relic abundance as function of DM mass for modified axion inflation for different $p$ values. The dashed band represents the observed relic abundance in the universe. (Right panel) Relic satisfying parameter space (magenta) in $\lambda_\chi-m_\chi$ plane when $\xi_\chi$ is varied in the range of 0.1-1 considering $p=3$ and $\Gamma=10^{12}$ GeV.}
\label{fig:extra1}
\end{figure}

\begin{figure}
\includegraphics[height=7cm,width=8cm]{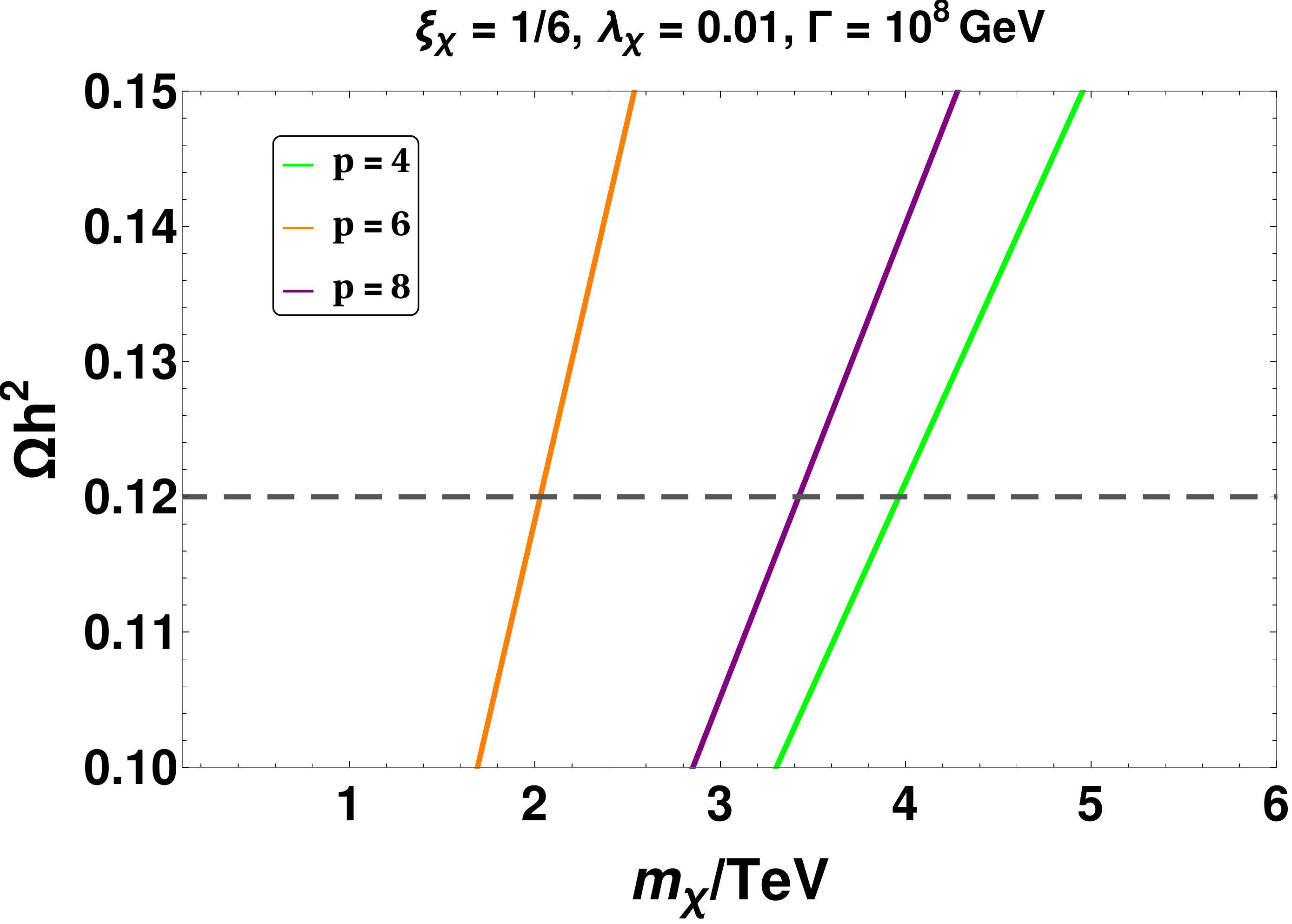}~~
\includegraphics[height=6.5cm,width=8.5cm]{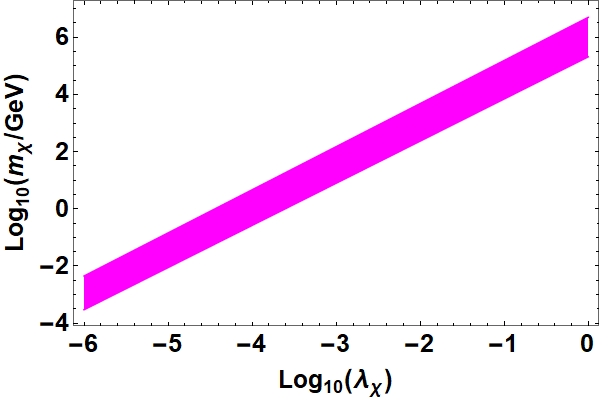}
\caption{(Left panel) DM relic abundance as function of DM mass for hilltop inflation for different $p$ values. The dashed band represents the observed relic abundance in the universe. (Right panel) Relic satisfying parameter space (magenta) in $\lambda_\chi-m_\chi$ plane when $\xi_\chi$ is varied in the range of 0.1-1 considering $p=4$ and $\Gamma=10^{8}$ GeV.}
\label{fig:extra2}
\end{figure}
Next we turn our attention to hilltop inflation \cite{Boubekeur:2005zm}. This is a low scale inflation model where the inflaton rolls from zero value. The generic form of the hilltop potential is \cite{Antusch:2019qrr}
\begin{align}
V(\phi)=V_0\left(1-\frac{\phi^p}{v_\phi^p}\right)^2 {~~~~\rm with~~~~ }p\geq 2 {\rm ~~~~and~ even}.
\end{align}
The potential has a plateau around $\phi\sim 0$, as well as two
minima at $\phi = \pm v_\phi$. The $V_0$ represents the energy density during inflation. This particular inflation model is consistent with the Planck 2018 data.
For example, with $p=4, (V_0)^{1/4}=1.1 \times 10^{14} ~{\rm ~GeV}, v_\phi = 3.9\times 10^{17}$ GeV the spectral index and tensor are estimated to be 0.967 and $10^{-10}$ respectively.
 
Since the natural inflation is a high scale inflation, we did not find much difference in the order of DM mass to satisfy the correct relic abundance compared to the non-minimal quartic inflation. In left panel of Fig. \ref{fig:extra1}, we present the relic abundance as function of mass of the DM considering the fixed values of the other relevant parameters.  We obtain $m_\chi\sim 17$ MeV (corresponding to $p=3$), when $\xi=1/6,\lambda_\chi=0.01$ and $\Gamma=10^{12}$ GeV to be consistent with the observed relic limit. We have also shown the relic contours for different $p$ values and estimated the required DM mass to satisfy the relic bound. The right panel of Fig. \ref{fig:extra1} shows the relic satisfied region in $\lambda_\chi-m_\chi$ plane considering $p=3$ and $\Gamma=10^{12}$ GeV. The $\xi_\chi$ has been varied in the range 0.1-1.

On the other hand, the hilltop inflation is a kind of low scale inflation with the order of Hubble parameter during inflation around $10^9$ GeV. Since the Hubble parameter is substantially small here compared to the other inflation models we discuss, the initial amplitude of DM oscillation remains small as well. Hence a larger DM mass is required to attain the correct order of relic. In left panel of Fig. \ref{fig:extra2}, we have shown the relic abundance as function of DM mass considering $\xi_\chi=1/6$, $\lambda_\chi=0.01$ and $\Gamma=10^8$ GeV. We notice that here the relic abundance is satisfied at a comparatively larger DM mass $\sim \mathcal{O}(1)$ TeV. Right panel of Fig. \ref{fig:extra2} exhibits the relic satisfied region in $\lambda_\chi-m_\chi$ plane considering $p=4$ and $\Gamma=10^{8}$ GeV. The $\xi_\chi$ has been varied in the range 0.1-1. In this figure the blue and red coloured regions do not appear since the corresponding bounds earlier defined are very relaxed due to smallness of the Hubble parameter (see equation \eqref{DMmasslb} and equation \eqref{eqn:lambdabound}).

\section{Conclusion}
\label{sec5}

To summarise, we provide a concrete particle physics framework to assimilate inflation, dark matter and Majorana neutrino mass with fate of each of these entities determined by gravitational effects. It is known that a non-minimal coupling of the inflaton with gravity can rescue the quartic chaotic inflation model. The light neutrino mass in the model is generated owing to the presence of Planck suppressed operators involving the inflaton which violate global lepton number symmetry explicitly, possibly by quantum gravity effects.  Notably, the main focus of the paper is to examine the DM production through classical misalignment mechanism in the non-minimal inflation framework. While gravitational misalignment mechanism as a source of DM production was discussed in earlier works, our work is a generalisation of these in the following aspects namely, (i) We consider a general class of inflation model with non-minimal coupling to gravity and  pursue the outcomes of the Palatini as well as Metric formalism of gravity in a detailed manner; (ii) The appearance of conformal factors in the coupled Boltzmann equations to solve for DM relic, resulted a different phenomenology compared to earlier works;  (iii) We also talk about the sensitivity of DM relic to inflationary parameter space, which was not indicated in the earlier works. In order to investigate this parameter dependence further, we generalise the study to two other inflationary models namely hilltop and natural inflation and found different DM mass range in one of these cases; (iv) On top of that, 
we build a well defined particle physics model which is able to accomodate light neutrino mass where the inflaton field is responsible for generating Majorana neutrino masses via type I seesaw mechanism, as mentioned above.

We have done a complete numerical exercise by solving the set of coupled Boltzmann equations relevant for DM, inflaton field as well as the SM radiation in each of the cases mentioned above instead of performing an approximate analytical analysis. We point out the consistent parameter spaces wherever applicable and highlight the new findings of the current analysis in the light of earlier works in this direction. In particular we find that all the three sectors namely inflation, dark matter and neutrino are essentially connected and any deviation of the parameters in a particular sector leads to different dynamics on another sector. As an example, it is observed that a change in magnitude of the non-minimal coupling of the inflaton results in different DM mass to obey the correct relic bound. Similar correlation also holds true in case of natural inflation and hilltop inflation models which are still consistent with Planck 2018 data. While we did not find much difference in natural inflation from what was obtained in non-minimal quartic inflation, we observe that hilltop inflation can allow comparatively heavier DM as well.

In a case where right handed neutrinos acquire masses from the VEV of the inflaton field, their phenomenology could depend crucially on the inflationary details as well as the reheat temperature. Since the present framework also explores such kind of connection, the right handed neutrinos can be useful in generating baryon asymmetry of the universe via leptogenesis \cite{Fukugita:1986hr}. Now if right handed neutrino mass turns out to be smaller than the reheat temperature one can realise thermal leptogenesis which is almost insensitive to the inflationary details. The opposite mass regime for the right handed neutrinos could lead to non-thermal leptogenesis scenario. Such a scenario was discussed recently in the context of non-minimal quartic inflation where, similar to the present proposal, the VEV of the inflaton field gives masses to the right handed neutrinos \cite{Borah:2020wyc}. Since we have a coupled system of right handed neutrinos, DM along with the inflaton with the last two fields coupled to gravity non-minimally, it will be worth exploring the dynamics of non-thermal leptogenesis in the present set up. We leave such a detailed study to an upcoming work.

\acknowledgements
DB acknowledges the support from Early Career Research Award from DST-SERB, Government of India (reference number: ECR/2017/001873). AKS is supported by  a post-doctoral fellowship at PRL, Ahmedabad.

%\bibliographystyle{JHEP}
%\bibliography{ref_db}

\providecommand{\href}[2]{#2}\begingroup\raggedright\endgroup

\end{document}